\documentclass[12pt]{article}
\pdfoutput=1
\usepackage[totalwidth=480pt,totalheight=640pt]{geometry}
\usepackage[centertags]{amsmath}
\usepackage[square, comma, sort&compress,numbers]{natbib}
\usepackage{array,multirow}
\usepackage{amssymb}
\usepackage{graphicx}
\usepackage{caption}
\usepackage{subcaption}
\usepackage{color}
\usepackage{setspace}
\usepackage{comment}
\usepackage{mathrsfs}

\usepackage{epsfig}
\usepackage{empheq}
\usepackage{enumitem}

\def\half{{\fr{1}{2}}}

\def\lb{\bar{\lambda}}

\def\Or[#1]{{\text{O}}\left({#1}\right)}
\def\dotl[#1,#2]{\left\langle #1, #2 \right\rangle}
\def\dotlb[#1,#2]{[ #1, #2 ]}
\def\dotp[#1,#2]{(#1) \cdot (#2)}
\def\aff[#1,#2]{\hat{#1}(#2)}
\def\n4sym{{\cal N}=4 SYM}
\def\>{\rangle}
\def\<{\langle}
\def\weight[#1,#2,#3]{\{(#1),#2,#3\}}
\def\ads[#1]{$\text{AdS}_{#1}$}

\newcommand{\ba}{\begin{eqnarray}}
\newcommand{\ea}{\end{eqnarray}}
\newcommand{\be}{\begin{equation}}
\newcommand{\ee}{\end{equation}}
\newcommand{\benn}{\begin{equation*}}
\newcommand{\eenn}{\end{equation*}}
\newcommand{\bi}{\begin{itemize}}
\newcommand{\ei}{\end{itemize}}

\newcommand{\Ccal}{{\cal C}}
\newcommand{\Hcal}{{\cal H}}

\newcommand{\Lcal}{{\cal L}}
\newcommand{\Mcal}{{\cal M}}

\newcommand{\Ocal}{{\cal O}}

\newcommand\oo\infty
\newcommand\s\sigma
\newcommand\de\delta
\newcommand\De\Delta
\newcommand{\ep}{\varepsilon}
\newcommand{\p}{\partial}
\newcommand\f\phi
\newcommand\g\gamma
\newcommand\x\times

\newcommand{\ra}{\rightarrow}

\newcommand{\fr}{\frac}

\newcommand{\comm}[2]{[#1,#2]}

\newcommand\G{\Gamma}

\newcommand{\even}{\textrm{even}}
\newcommand{\odd}{\textrm{odd}}
\newcommand{\gap}{\textrm{gap}}
\newcommand{\KL}{K\"{a}ll\'{e}n-Lehmann }

\newcommand{\CFT}{\textrm{CFT}}
\newcommand{\Cmax}{{\cal C}_{\max}}
\newcommand{\Dmax}{\Delta_{\max}}

\newcommand{\LambdaIR}{\Lambda_{\textrm{IR}}}
\newcommand{\phivec}{\vec{\phi}^{\,2}}

\newcommand{\Lvec}{\boldsymbol{\ell}}

\newcommand{\sigvec}{\boldsymbol{\sigma}}

\newcommand{\kvec}{\boldsymbol{k}}

\newcommand{\norder}[1]{%
  {:\mathrel{\mspace{1mu}#1\mspace{1mu}}:}%
}

\newcommand{\lambar}{\bar{\lambda}}

\newcommand\lrpar{\raise .8ex\hbox{$^\leftrightarrow$} \hspace{-9pt}
\partial}
\newcommand\lpar{\raise .8ex\hbox{$^\leftarrow$} \hspace{-9pt}
\partial}
\newcommand\rpar{\raise .8ex\hbox{$^\rightarrow$} \hspace{-9pt}
\partial}
\newcommand\lrd{\raise .8ex\hbox{$^\leftrightarrow$} \hspace{-9pt}
\nabla}

\renewcommand{\hat}{\widehat}

\usepackage{hyperref}

\numberwithin{equation}{section}

\begin{document}

\begin{titlepage}

\begin{center}
\vspace{1cm}

{\Large \bf RG Flow from $\phi^4$ Theory to the 2D Ising Model}

\vspace{0.8cm}

\normalsize
\bf{Nikhil Anand$^1$, Vincent X.\ Genest$^2$, Emanuel Katz$^3$,\\\vspace{1mm} Zuhair U.\ Khandker$^{3,4}$, Matthew T.\ Walters$^3$}
\normalsize

\vspace{.5cm}

{\it $^1$ Department of Physics and Astronomy, Johns Hopkins University, Baltimore, MD 21218} \\
{\it $^2$ Department of Mathematics, Massachusetts Institute of Technology, Cambridge, MA 02139} \\
{\it $^3$ Department of Physics, Boston University, Boston, MA 02215}\\
{\it $^4$ Department of Physics, University of Illinois, Urbana, IL 61801}\\

\end{center}

\vspace{1cm}

\begin{abstract}

We study 1+1 dimensional $\phi^4$ theory using the recently proposed method of conformal truncation. Starting in the UV CFT of free field theory, we construct a complete basis of states with definite conformal Casimir, $\mathcal{C}$. We use these states to express the Hamiltonian of the full interacting theory in lightcone quantization. After truncating to states with $\mathcal{C} \leq \mathcal{C}_{\max}$, we numerically diagonalize the Hamiltonian at strong coupling and study the resulting IR dynamics. We compute non-perturbative spectral densities of several local operators, which are equivalent to real-time, infinite-volume correlation functions. These spectral densities, which include the Zamolodchikov $C$-function along the full RG flow, are calculable at any value of the coupling. Near criticality, our numerical results reproduce correlation functions in the 2D Ising model.

\end{abstract}

\bigskip

\end{titlepage}

\hypersetup{pageanchor=false}

\tableofcontents

%%%%%%%%%%%%%%%%%%%%%%%%%%%%%%%%%%%%%%%%%%%%%%%%%%%%%%%%%%%%%%%%%%%%%%%%%%%%%
%%%%%%%%%%%%%%%%%%%%%%%%%%%%%%%%%%%%%%%%%%%%%%%%%%%%%%%%%%%%%%%%%%%%%%%%%%%%%

\section{Introduction}
\label{sec:Intro}

The language of quantum field theory underpins our understanding of a vast array of physical phenomena. For strongly-coupled QFTs, however, we face a shortage of robust methods for calculating non-perturbative dynamics. In particular, apart from certain highly specialized examples, it is challenging in most methods to compute time-dependent observables, such as correlation functions of local operators or the wavefunctions of states. In \cite{Truncation}, we presented a new framework, which we called \emph{conformal truncation}, for computing real-time, infinite-volume observables in a non-perturbative QFT in any number of spacetime dimensions, given information about the UV conformal field theory from which it originates. In that work, the method was only tested in examples with a perturbative or large-$N$ expansion. The goal of the present work is to apply conformal truncation in a truly non-perturbative setting, and in so doing, to lay the groundwork for using this method to study dynamics in general QFTs.

Conformal truncation is a particular implementation of a more general approach known as Hamiltonian truncation (for a recent review, see \cite{James:2017cpc}). The basic strategy is to discretize the QFT Hilbert space in some way and then truncate it to a finite-dimensional subspace. The resulting truncated Hamiltonian can be diagonalized numerically, yielding an approximation to the true QFT spectrum. More importantly, we also obtain an approximation to the actual Hamiltonian eigenstates, which can be used to compute dynamical observables. The heart of any Hamiltonian truncation method is the discretization prescription, since it determines which symmetries are preserved under truncation, how efficiently IR degrees of freedom are captured, and, ultimately, which physical observables are deliverable. 

The method proposed in \cite{Truncation} uses conformal symmetry as the organizing principle for truncation. One starts by viewing the QFT in question as arising from a deformed UV CFT. A basis for the QFT Hilbert space is constructed in terms of UV fields and organized into representations of the conformal group, characterized by the quadratic Casimir eigenvalue $\Ccal$. One truncates the basis by specifying some maximum Casimir eigenvalue $\Cmax$ and only keeping states below this threshold. In this basis, matrix elements of the Hamiltonian are simply related to OPE coefficients of the UV CFT. Although the basis and Hamiltonian are constructed in the UV, after diagonalization, they describe the entire RG flow of the QFT. In this way, one is using CFT data to study QFT dynamics. 

A key feature of conformal truncation is that one can use it to compute real-time, continuum correlation functions. This is largely because the method avoids spacetime compactification or latticization. For two-point functions, one can compute the associated \KL spectral densities, $\rho(\mu)$, which encode the decomposition of these correlators in terms of mass eigenstates,
\be
\<\Ocal(x) \Ocal(0)\> = \int d\mu^2 \, \rho_\Ocal(\mu) \int \fr{d^dp}{(2\pi)^d} \, e^{-ip\cdot x} \, \theta(p_0) \, (2\pi) \de(p^2 - \mu^2).
\label{eq:SpecDef}
\ee
In \cite{Truncation}, we confirmed that conformal truncation indeed correctly reproduces known spectral densities in a large-$N$ example. Our goal here is to now use conformal truncation to compute fully non-perturbative spectral densities.

To have an independent check of our numerical results, we would like to study a QFT with two properties: (i) it originates from a UV CFT where we know operator dimensions and OPE coefficients so that we can construct the Hamiltonian, and (ii) it has some regime that is strongly-coupled, but with known analytic expressions for correlation functions that we can compare with our conformal truncation results. One QFT that satisfies these requirements is 1+1 dimensional $\phi^4$ theory, which can be viewed as the free massless CFT deformed by a mass term and quartic coupling, leading to the full Lagrangian\footnote{The operators in this Lagrangian are normal-ordered, but we have suppressed the typical notation, $\norder{\Ocal}$, with the understanding that \emph{all} local operators in this work are to be normal-ordered.} 
\be
\Lcal = \Lcal_{\CFT} + \de \Lcal = \half \p_\mu \phi \p^\mu \phi - \half m^2 \phi^2 - \fr{1}{4!} \lambda \phi^4.
\label{eq:L}
\ee
The dynamics of this theory are controlled by the dimensionless parameter
\benn
\lambar \equiv \frac{\lambda}{m^2}.
\eenn
Using conformal truncation, we can compute spectral densities for \emph{any} $\lambar$. To the best of our knowledge, this is the first calculation of non-perturbative spectral densities in 2D $\phi^4$ theory.

For some critical value $\lambar_*$, the mass gap closes and the theory flows to a non-trivial IR fixed point in the same universality class as the critical 2D Ising model, a theory for which many exact results are known. We can thus test conformal truncation in a strongly-coupled setting by comparing the IR behavior of our resulting spectral densities in the vicinity of the critical point to the known analytic expressions for the Ising model.

We focus specifically on the local operators $\phi^n$ and the stress-energy tensor $T_{\mu\nu}$. Our results for the spectral densities of these operators can be summarized as follows:
\bi[leftmargin=*]
\itemsep0em
\item We verify explicitly that $\phi^4$ theory at $\lambar_*$ flows to a non-trivial CFT. Specifically, we compute the spectral density of the trace of the stress tensor, $T^\mu_{\phantom{\mu}\mu}$, and confirm that near criticality it reproduces the 2D Ising prediction in the IR, vanishing as $\lambar\ra\lambar_*$. (Figure~\ref{fig:TraceZoomIn})
\item We demonstrate universality in the IR behavior of $\phi^n$ correlators near criticality. In particular, we find that the spectral densities of the even operators $\phi^{2n}$ all match the Ising model prediction for $\ep$, while the odd operators $\phi^{2n-1}$ match the prediction for $\s$. (Figures~\ref{fig:PhiNConvergence} and~\ref{fig:PhiNZoomOut})
\item We compute the Zamolodchikov $C$-function along the full RG flow. We find that it decreases monotonically from the free central charge $c_{\textrm{UV}} = 1$, transitioning to the strongly-coupled IR at a scale set roughly by the coupling $\fr{\lambda}{4\pi}$. Near criticality, the IR behavior agrees with the prediction from the Ising model. (Figure~\ref{fig:CentralChargeConvergence})
\ei
It is worth emphasizing that our numerical results for the spectral densities describe the \emph{entire} RG flow, not just the IR regime described by the Ising model. In addition, we can use conformal truncation to compute dynamical observables at any value of the coupling, not just the narrow range near $\lambar_*$. We merely choose to focus on the vicinity of the critical point in this work in order to test our framework against analytic results.

There have been many previous applications of Hamiltonian truncation methods to two-dimensional $\phi^4$ theory \cite{Harindranath:1987db,Harindranath:1988zt,Burkardt:2016ffk,Lee:2000ac,Sugihara:2004qr,Schaich:2009jk,Milsted:2013rxa,Bosetti:2015lsa,Rychkov:2014eea,Rychkov:2015vap,Rozowsky:2000gy,Chakrabarti:2003ha,Chakrabarti:2003tc,Chakrabarti:2005zy,Coser:2014lla,Elias-Miro:2015bqk,Bajnok:2015bgw,Elliott:2014fsa,Chabysheva:2016ehd,Christensen:2016naf}. In particular, Burkardt et al.~\cite{Burkardt:2016ffk} have proposed using a Fock space basis of symmetric polynomials which in fact match the Casimir eigenstates we use to construct our basis. However, our approach differs somewhat from theirs in practice, as we truncate our basis solely according to Casimir eigenvalue, keeping higher-particle states which they neglect. In addition, we use the conformal structure of the UV theory to simplify the construction of the basis, allowing us to significantly increase the number of states and compute full spectral densities.

Looking forward, conformal truncation can be applied to deformations of more general CFTs, in any number of dimensions, provided we have sufficient knowledge of scaling dimensions and OPE coefficients to construct the Hamiltonian. Our results for $\phi^4$ theory thus provide a first step toward using this method to study a variety of strongly-coupled dynamics.

The outline of the paper is as follows. In section~\ref{sec:Model}, we briefly review the general framework of conformal truncation and discuss its application to 1+1 dimensional scalar field theory. In section~\ref{sec:SanityChecks} we perform some simple consistency checks, numerically reproducing several free field theory spectral densities and then verifying the constraints imposed by the equation of motion and conservation of the stress-energy tensor. In section~\ref{sec:Coupling}, we proceed to strong coupling, studying the behavior of the low-mass spectrum as a function of the coupling $\lambar$ in order to determine the point at which the mass gap closes. We then extrapolate the truncated results to determine a prediction for the critical coupling, $\lambar_*$, which we compare to previous results in the literature. In section~\ref{sec:Ising}, we compute spectral densities in the vicinity of the critical point, comparing the results to analytic predictions from the Ising model. We conclude and discuss future directions in section~\ref{sec:Discussion}, while several appendices contain details of our methods.

%%%%%%%%%%%%%%%%%%%%%%%%%%%%%%%%%%%%%%%%%%%%%%%%%%%%%%%%%%%%%%%%%%%%%%%%%%%%%
%%%%%%%%%%%%%%%%%%%%%%%%%%%%%%%%%%%%%%%%%%%%%%%%%%%%%%%%%%%%%%%%%%%%%%%%%%%%%

\section{Conformal Truncation and Scalar Field Theory}
\label{sec:Model}

The goal of this work is to use conformal truncation to study the RG flow of 1+1 dimensional $\phi^4$ theory, given by the Lagrangian in eq.~(\ref{eq:L}), to the 2D Ising model. In this section, we introduce all of the necessary ingredients to accomplish this task. We first review the overall approach of conformal truncation and then discuss the details of applying this method to the specific UV CFT of 2D free scalar field theory. Finally, we briefly review spectral densities, which are our main dynamical observable.

%%%%%%%%%%%%%%%%%%%%%%%%%%%%%%%%%%%%%%%%%%%%%%%%%%%%%%%%%%%%%%%%%%%%%%%%%%%%%

\subsection{Review of Conformal Truncation}

Conformal truncation is a method for using CFT data to numerically study the IR dynamics of more general QFTs. This method can be applied to any theory that can be described as an RG flow originating from some UV CFT deformed by one or more relevant operators,
\be
S = S_{\CFT} - \lambda \int d^dx \, \Ocal_R(x).
\ee
Following the approach presented in \cite{Truncation}, a useful basis for the Hilbert space of this theory consists of UV eigenstates of the quadratic Casimir of the conformal group,
\be
|\Ccal,\vec{P},\mu\> \equiv \int d^dx \, e^{-iP\cdot x} \Ocal(x)|0\>,
\label{eq:basis}
\ee
where $\mu^2 \equiv P^2$. These basis states are created by primary operators\footnote{In this work, ``primary'' refers to any operator which is primary with respect to the global conformal group $SO(d,2)$ and thus annihilated by the special conformal generators ($\comm{K_\mu}{\Ocal(0)} = 0$). In 2D, this includes operators which are often referred to as ``quasi-primary'' or ``global primary'' in the literature.} in the original CFT, and are characterized by their Casimir eigenvalue, spatial momentum, and invariant mass (suppressing other possible quantum numbers like the spin $\ell$). 

The strategy of conformal truncation is to restrict the Hilbert space to the subspace spanned by states with Casimir eigenvalue $\Ccal \leq \Cmax$. The full Hamiltonian (CFT + deformation), when restricted to this subspace, can be diagonalized numerically, yielding an approximation to the true spectrum of the IR QFT.

To define the Hamiltonian, we first need to choose a quantization scheme. As discussed in \cite{Truncation}, we work in lightcone quantization, with the Hilbert space defined on slices of constant lightcone ``time'' $x^+ \equiv \fr{1}{\sqrt{2}}(t+x)$. We thus need to compute matrix elements for the associated lightcone Hamiltonian
\be
P_+ = P_+^{(\CFT)} + \lambda \int d^{d-1}\vec{x} \, \Ocal_R(x^+=0,\vec{x}).
\ee
By construction, our basis is built from eigenstates of the CFT Hamiltonian, so we only need to compute matrix elements associated with the relevant deformation. These matrix elements are simply Fourier transforms of three-point functions in the original UV CFT,
\be
\<\Ccal,\vec{P},\mu| \de P_+ |\Ccal',\vec{P}',\mu'\> = \lambda \int d^d x \, d^{d-1}\vec{y} \, d^dz \, e^{i(P\cdot x - P'\cdot z)} \<\Ocal(x) \Ocal_R(y) \Ocal'(z)\>.
\label{eq:MatrixDef}
\ee
We thus only need data from the UV fixed point to study the full RG flow: the spectrum of local operators gives us a complete basis, while the OPE coefficients give us the Hamiltonian matrix elements.

%%%%%%%%%%%%%%%%%%%%%%%%%%%%%%%%%%%%%%%%%%%%%%%%%%%%%%%%%%%%%%%%%%%%%%%%%%%%%

\subsection{Conformal Basis for 2D Scalar Fields}

Our starting point is the 2D free massless scalar in the UV. To apply conformal truncation, we need to first construct the complete set of primary operators built from the scalar field $\phi$.\footnote{This basis was originally considered in~\cite{Katz:2013qua,Katz:2014uoa}, though with the separate goal of studying bound states in 2D QCD.} This process is more subtle than in higher dimensions, because in 2D $\phi$ is \emph{not} a primary operator. We can see this by looking at its two-point function, which is logarithmically divergent,
\be
\<\phi(x) \phi(0)\> = \fr{-1}{2\pi} \log|x|.
\label{eq:phi2pt}
\ee
In order to construct well-defined primary operators, we must instead use the ``building blocks''
\benn
\p_-\phi, \quad \p_+\phi, \quad e^{i\alpha\phi}.
\eenn
However, for the purposes of conformal truncation we do not need the latter two, as we now explain. 

Consider $\p_+\phi$. From the equations of motion, we see that $\p_{\pm}\phi$ are purely left-moving and right-moving modes, respectively,
\be
\p^2 \phi = \p_-(\p_+\phi) = \p_+(\p_-\phi) = 0.
\ee
The left-moving operator $\p_+\phi$ thus creates particles with zero lightcone momentum $P_-$. Because we are working in lightcone quantization, these left-moving states are non-dynamical and can be integrated out, setting $\p_+\phi =0$ \cite{Chao:1993bm}.

Now consider the vertex operators $e^{i\alpha\phi}$, parameterized by the variable $\alpha$. Because of the logarithmic divergence in eq.~\eqref{eq:phi2pt}, these operators require the introduction of an IR scale $R$,
\be
\<e^{i\alpha\phi(x)} e^{-i\alpha\phi(0)}\> = \sum_n \fr{\alpha^{2n}}{(n!)^2} \<\phi^n(x) \phi^n(0)\> = \sum_n \fr{\alpha^{2n}}{n!(2\pi)^n} \log^n \left|\fr{R}{x}\right| = \left( \fr{R}{x} \right)^{\fr{\alpha^2}{2\pi}}.
\ee
This IR scale can be absorbed into a redefinition of $e^{i\alpha\phi}$, yielding a well-defined set of primary operators with scaling dimensions $\De_\alpha = \fr{\alpha^2}{4\pi}$. However, once we deform the UV CFT by adding the mass term
\benn
\de \Lcal = -\half m^2 \phi^2,
\eenn
the resulting Hamiltonian matrix elements for these vertex operators depend on the IR scale, diverging as $R \ra \infty$. These divergences ``lift'' the vertex operators from the theory, such that they have no overlap with the physical low-energy states. This behavior is unsurprising, as vertex operators cease to be independent degrees of freedom in the massive theory.

Consequently, we can ignore both left-moving and vertex operators.\footnote{The removal of vertex operators and the restriction to states built from $\p_- \phi$ is quite similar to the construction of the ``Dirichlet basis'' discussed in \cite{Truncation}.} Thus our basis consists of primary operators built only from the right-moving mode $\p_-\phi$, with the general form
\be
\boxed{\Ocal(x) = \sum_{\kvec} C^\Ocal_{\kvec} \, \p_-^{k_1} \phi(x) \cdots \p_-^{k_n} \phi(x),}
\ee
for some coefficients $C^\Ocal_{\kvec}$ that need to be determined. The method for constructing these primary operators is discussed in appendix~\ref{app:Basis} and will be presented in more detail in~\cite{FutureUs}. Because these operators only consist of right-moving modes, their associated conformal Casimir eigenvalues are completely fixed by their scaling dimensions,
\be
\Ccal = \De(\De-2) + \ell^2 = 2\De(\De-1).
\ee
Setting a maximum Casimir eigenvalue, $\Cmax$, is thus equivalent to setting a maximum scaling dimension, $\Dmax$.

The right-moving operators are all annihilated by the original CFT Hamiltonian,
\be
\comm{P_+^{(\CFT)}}{\Ocal(x)} = 0.
\ee
This means that \emph{all} states built from these primary operators have zero invariant mass $P^2$. Thus for the 2D free scalar, the conformal truncation basis states in eq.~(\ref{eq:basis}) take the more restricted form
\be
|\Ccal,P_-,\mu=0\> \equiv \int dx^- \, e^{-iP_- x^-} \Ocal(x^-)|0\>.
\label{eq:2Dbasis}
\ee

Unlike in higher dimensions, where each primary operator defines a continuum of Casimir eigenstates, parameterized by the invariant mass $\mu$, each 2D operator $\Ocal$ only defines a \emph{single} basis state. For a given $\Cmax$, the number of states in our basis is therefore given by the number of primary operators with Casimir eigenvalue below that threshold. It is important to note that this significant reduction of the basis is specific to 2D free field theory (or more generally, 2D theories built from conserved currents). In other CFTs, primary operators are \emph{not} annihilated by $P_+$, leaving the invariant mass $\mu$ as a continuous parameter defining a multiplet of Casimir eigenstates for each operator.

After constructing the basis, the next step is to work out Hamiltonian matrix elements. Since $P_+^{(\CFT)}$ vanishes in our basis, the full lightcone Hamiltonian only has contributions from the relevant deformations,
\be
\boxed{P_+ = \int dx^- \left( \half m^2 \phi^2 + \fr{1}{4!} \lambda \phi^4 \right).}
\ee
We can compute the Hamiltonian matrix elements by Fourier transforming three-point functions involving $\phi^2$ and $\phi^4$, following eq.~\eqref{eq:MatrixDef}. Because these relevant deformations are \emph{not} primary operators, their three-point functions are not simply a universal kinematic factor multiplied by an overall OPE coefficient. Fortunately, their correlation functions can all easily be computed via Wick contractions. The resulting matrix elements are presented in appendix~\ref{app:Matrix}.

%%%%%%%%%%%%%%%%%%%%%%%%%%%%%%%%%%%%%%%%%%%%%%%%%%%%%%%%%%%%%%%%%%%%%%%%%%%%%

\subsection{Review of Spectral Densities}

After we have truncated the basis to some $\Cmax$ and computed the associated Hamiltonian matrix elements, we can construct the invariant mass operator
\be
M^2 = 2P_+ P_-.
\ee
Because our basis consists of $P_-$ eigenstates, diagonalizing this Lorentz invariant operator is actually equivalent to diagonalizing the lightcone Hamiltonian $P_+$.

The mass eigenvalues that result from diagonalizing $M^2$ are an approximation to the spectrum of the IR QFT. However, in addition to the eigenvalues, we also obtain the associated eigenstates $|\mu_i\>$, which we can use to compute dynamical IR observables. One natural and important observable for us to study is the spectral density of any local operator $\Ocal(x)$,
\be
\rho_\Ocal(\mu) \equiv \sum_i |\<\Ocal(0)|\mu_i\>|^2 \, \de(\mu^2 - \mu_i^2).
\label{eq:rho}
\ee
As shown in eq.~\eqref{eq:SpecDef}, spectral densities encode the same information as real-time, infinite-volume correlation functions. For presenting results, it will be more convenient to show the integrated spectral density,
\be
I_\Ocal(\mu) \equiv \int_0^{\mu^2} d\mu^{\prime\,2} \, \rho_\Ocal(\mu') = \sum_{\mu_i \leq \mu} |\<\Ocal(0)|\mu_i\>|^2,
\label{eq:I}
\ee
which contains the same dynamical information as the spectral density.

%%%%%%%%%%%%%%%%%%%%%%%%%%%%%%%%%%%%%%%%%%%%%%%%%%%%%%%%%%%%%%%%%%%%%%%%%%%%%
%%%%%%%%%%%%%%%%%%%%%%%%%%%%%%%%%%%%%%%%%%%%%%%%%%%%%%%%%%%%%%%%%%%%%%%%%%%%%

\section{Sanity Checks}
\label{sec:SanityChecks}

In this section, we perform two consistency checks of our conformal truncation method. First, we consider the free field theory limit, $\lambar=0$, and verify that our numerical results for $\phi^n$ spectral densities match the theoretical predictions. Second, we confirm that the equation of motion and the stress-energy tensor Ward identity are satisfied identically in our framework for any $\lambar$, even after truncation.

%%%%%%%%%%%%%%%%%%%%%%%%%%%%%%%%%%%%%%%%%%%%%%%%%%%%%%%%%%%%%%%%%%%%%%%%%%%%%

\subsection{Spectral Densities in Free Field Theory}

Here we consider free massive field theory, obtained by setting $\lambar=0$. In this limit, Hamiltonian matrix elements are diagonal with respect to particle number, which means that we can consider each $n$-particle sector independently. For each sector, we truncate the basis to some $\Dmax$ (or equivalently $\Cmax$), diagonalize the lightcone Hamiltonian, and use the resulting approximate mass eigenstates to compute the spectral density of the corresponding scalar operator $\phi^n$.

\begin{figure}[t!]
\begin{center}
\includegraphics[width=\textwidth]{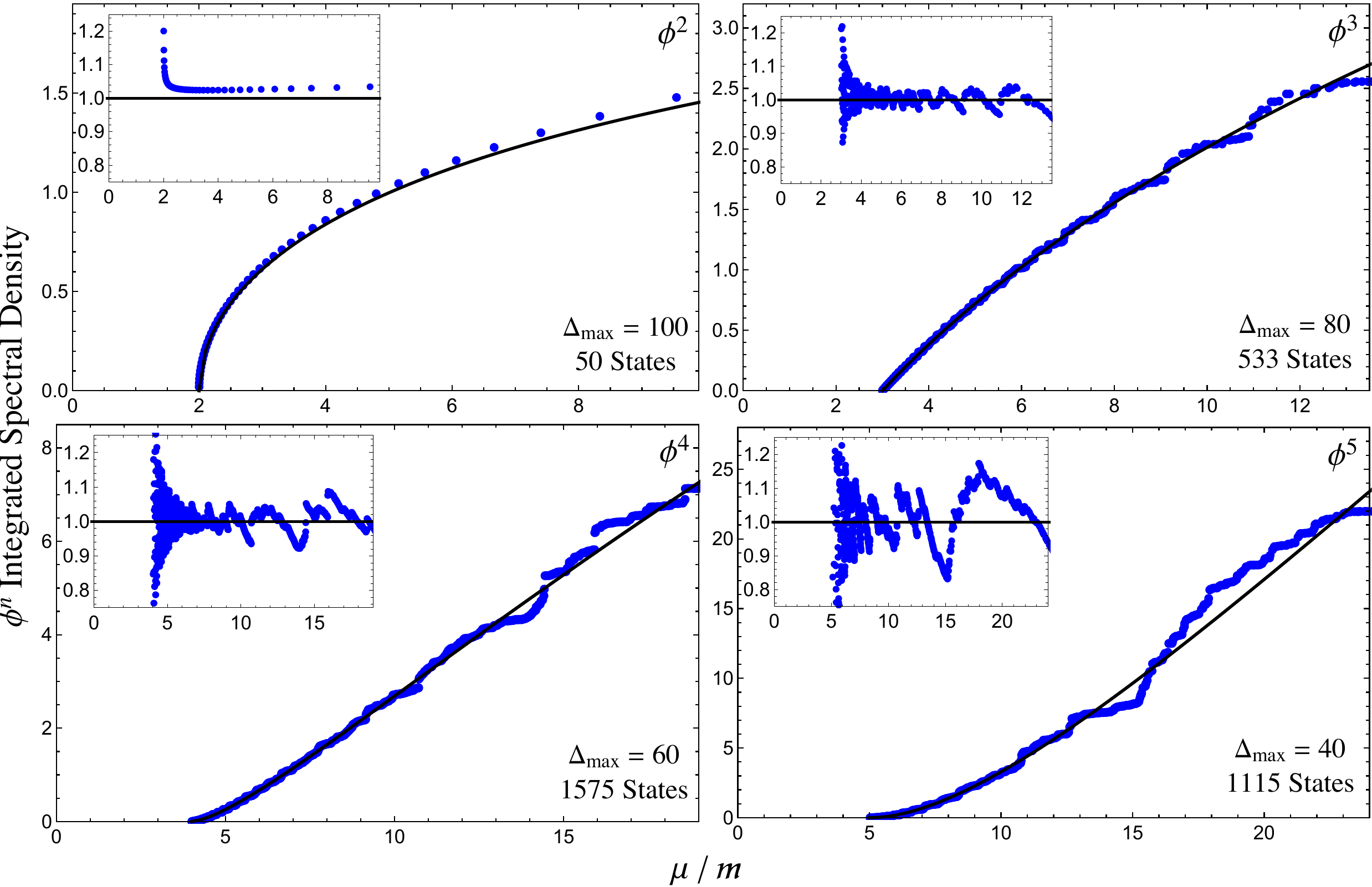}
\caption{Integrated spectral densities for $\phi^2$ (upper left), $\phi^3$ (upper right), $\phi^4$ (lower left), and $\phi^5$ (lower right) in massive free field theory ($\lambar=0$), both the raw value (main plot) and normalized by the theoretical prediction (inset). The conformal truncation results (blue dots) for each plot are computed using the $\Dmax$ shown, with the corresponding number of $n$-particle basis states, and compared to the theoretical prediction (black curve).}
\label{fig:FreeDensities} 
\end{center}
\end{figure}

As examples, figure~\ref{fig:FreeDensities} shows the integrated spectral densities for $\phi^2$, $\phi^3$, $\phi^4$, and $\phi^5$. In each plot, the blue dots are our conformal truncation results and the black line is the theoretical prediction, given by~\cite{Groote:1998ic}
\be
\begin{split}
\rho_{\phi^n}(\mu) = \frac{n!}{2^n \pi^{n+1}} &\int_0^\infty dr \, r K_0(\mu r) \\
& \qquad \times \Big[2 K_0(m r)^n - \big(K_0(m r)+i \pi I_0(m r)\big)^n - \big(K_0(m r)-i \pi I_0(m r)\big)^n  \Big],
\end{split}
\ee
where $I_0$ and $K_0$ are modified Bessel functions of the first and second kind.

The main plot shows the raw value for the integrated spectral density, while the inset shows the same result normalized by the prediction. For each plot, we also indicate the number of $n$-particle basis states for the corresponding choice of $\Dmax$. For example, for $\phi^2$ we set $\Dmax=100$, meaning we have kept all 2-particle states with $\De \leq 100$, which corresponds to a total of $50$ states.

As is evident from the figure, the conformal truncation results correctly reproduce the theoretical expectations for these spectral densities. Similar plots can also be made for $\phi^n$ with $n>5$. These plots serve as both a consistency check of our method, ensuring that our basis states and matrix elements have been constructed correctly, as well as a demonstration that our conformal truncation approach can be used to compute full correlation functions.

From the insets in figure~\ref{fig:FreeDensities}, we see that the numerical results agree with the full functional form of the spectral density to within a few percent over a wide range of $\mu$. The discrepancy slowly begins to increase in the UV, confirming that our basis of primary operators with low conformal Casimir predominantly overlaps with low-mass states \cite{Fitzpatrick:2013twa}. The discrepancy also grows rapidly near the IR threshold $\mu\approx nm$. This is due to the fact that we have truncated to a discrete basis, giving rise to an effective IR cutoff (see appendix~\ref{app:Decoupling}),
\be
\LambdaIR \sim \fr{m}{\Dmax}.
\ee
Increasing $\Dmax$ lowers this effective cutoff, improving our ability to resolve IR mass scales.

%%%%%%%%%%%%%%%%%%%%%%%%%%%%%%%%%%%%%%%%%%%%%%%%%%%%%%%%%%%%%%%%%%%%%%%%%%%%%

\subsection{Equation of Motion and Ward Identity}

In our framework, both the equation of motion (EOM) and the Ward identity for the stress-energy tensor can be phrased as constraints on certain matrix elements of the invariant mass operator $M^2$. It is convenient to specifically focus on the dynamical part of these matrix elements, $\Mcal_{\Ocal\Ocal'}$, with the overall momentum-conserving delta function removed, 
\be
\<\Ccal,P_-|M^2|\Ccal',P_-'\> \equiv 2P_- (2\pi) \de(P_- - P_-') \, \Mcal_{\Ocal\Ocal'}.
\ee

To derive the matrix element constraints imposed by the EOM, we start with the equation in operator form and act on the vacuum to obtain the relation
\be
M^2 \phi(0)|0\> = m^2 \phi(0)|0\> + \frac{1}{3!} \lambda \phi^3(0)|0\>.
\label{eq:EOM1}
\ee
We now act on both sides with an arbitrary basis state $\<\Ccal,P_-|$, obtaining the constraint
\be
\Mcal_{\Ocal,\p\phi} = m^2 \<\Ccal,P_-|\phi(0)\> + \frac{1}{3!} \lambda \<\Ccal,P_-|\phi^3(0)\>.
\label{eq:EOM2}
\ee
The left side of this equation is an $M^2$ matrix element mixing the one-particle state with a generic basis state created by any primary operator $\Ocal$. The EOM thus relates this matrix element to the overlap that the $\Ocal$ basis state has with $\phi$ and $\phi^3$. Using the matrix elements presented in appendix~\ref{app:Matrix}, it is straightforward to check that eq.~\eqref{eq:EOM2} indeed holds for any state in our basis. Since the EOM is satisfied at the level of individual matrix elements, it holds exactly for the resulting mass eigenstates, regardless of how we truncate the basis.

The EOM is a useful warmup for the stress-energy tensor Ward identity,
\be
P^\mu T_{\mu\nu} = P_+ T_{--} + P_- T_{+-} = 0.
\label{eq:Ward}
\ee
In 2D $\phi^4$ theory, the momentum generators are defined as 
\be
P_- \equiv \int dx^- \left(\p_- \phi \right)^2, \qquad P_+ \equiv \int dx^- \left( \half m^2 \phi^2 + \fr{1}{4!} \lambda \phi^4 \right).
\ee
Given these integral expressions for $P_\pm$, by the Noether construction one would na\"{i}vely expect the components $T_{--}$ and $T_{+-}$ to be given by the corresponding integrands. While this expectation is correct for $T_{--}$,\footnote{Note that our definition of $T_{--}$ differs from the standard one (in e.g.~\cite{YellowPages}) by a factor of $2\pi$.}
\be
T_{--} \equiv \left(\p_- \phi \right)^2,
\label{eq:T--}
\ee 
it is \emph{not} true for $T_{+-}$. This subtlety in defining the stress tensor arises from the fact that the scalar field $\phi$ is not a well-defined primary operator.

To see this concretely, consider the OPE of $T_{--}$ with a general scalar primary operator $\Ocal$ in any 2D CFT,
\be
T_{--}(x)\Ocal(y) \sim \frac{-\De_\Ocal}{4\pi (x^--y^-)^2} \Ocal(y) - \frac{1}{2\pi (x^--y^-)} \p_-\Ocal(y) + \cdots 
\ee
where the remaining terms in the expansion are not singular. For the operator $\phi^4$, however, we instead have the peculiar expansion
\be
T_{--}(x) \phi^4(y) \sim \frac{3}{4\pi^2(x^--y^-)^2} \phi^2(y) - \frac{1}{2\pi (x^--y^-)} \p_-\phi^4(y) + \cdots
\ee
Thus $\phi^4$ can give rise to $\phi^2$, such that the distinction between the two operators is muddied.

We can use the Ward identity to determine the correct form of $T_{+-}$. Using the OPE, one can check explicitly that eq.~\eqref{eq:Ward} requires
\be
\boxed{T_{+-} \equiv \frac{1}{2}m^2\phi^2 + \frac{1}{4!} \lambda \phi^4 + \frac{1}{16\pi} \lambda \phi^2.}
\label{eq:T+-}
\ee
While there is a discrepancy between this expression for $T_{+-}$ and the integrand of $P_+$, this appears to be an unavoidable pathology of 2D scalar field theory due to the fact that we have chosen to deform the UV CFT by an ill-defined operator.

Nevertheless, we can confirm that the expression for $T_{+-}$ above is correct by studying the matrix element constraints imposed by the Ward identity. Following the same procedure as the EOM, the Ward identity implies
\be
\Mcal_{\Ocal,(\p\phi)^2} = \sqrt{48\pi} \<\Ccal,P_-|T_{+-}(0)\>,
\ee
which constrains matrix elements involving the two-particle state created by $(\p_-\phi)^2$. Using the matrix elements in appendix~\ref{app:Matrix}, one can check that this constraint is only satisfied if we use the expression for $T_{+-}$ in eq.~\eqref{eq:T+-}. This consistency check is important, as we later use this expression to study the stress tensor spectral density in section~\ref{sec:Ising}.

%%%%%%%%%%%%%%%%%%%%%%%%%%%%%%%%%%%%%%%%%%%%%%%%%%%%%%%%%%%%%%%%%%%%%%%%%%%%%
%%%%%%%%%%%%%%%%%%%%%%%%%%%%%%%%%%%%%%%%%%%%%%%%%%%%%%%%%%%%%%%%%%%%%%%%%%%%%

\section{Critical Coupling for $\phi^4$ Theory}
\label{sec:Coupling}

In order to study the RG flow from scalar field theory to the 2D Ising model, we need to determine the critical coupling, $\lambar_*$. To do so, we scan over $\lambar$, diagonalizing the Hamiltonian for each value of the coupling to obtain the mass spectrum, and look for the following indicators of critical behavior:
\bi
\item {\bf Vanishing mass gap}. Since in lightcone quantization the vacuum is trivial \cite{Leutwyler:1970wn,Maskawa:1975ky}, the mass gap is simply the lowest mass eigenvalue. The critical coupling should therefore correspond to the point at which the lowest eigenvalue goes to zero. 
\item {\bf Continuous spectrum}. At weak coupling, the lowest eigenvalue corresponds to the one-particle state, which is separated from the two- and three-particle thresholds. At the critical coupling, not only should the lowest eigenvalue hit zero, but this spacing between eigenvalues should also vanish, providing an important consistency check that we have successfully tuned to the critical point.
\ei
In this section, we use these criteria to determine the value of the critical coupling. We study the mass spectrum as a function of $\lambar$ at various finite values for $\Dmax$ and then extrapolate the results to the limit $\Dmax\ra\infty$ to calculate $\lambar_*$. We then compare the value we obtain with previous results and briefly discuss the mapping between critical couplings in lightcone quantization with those in more standard equal-time quantization.

%%%%%%%%%%%%%%%%%%%%%%%%%%%%%%%%%%%%%%%%%%%%%%%%%%%%%%%%%%%%%%%%%%%%%%%%%%%%%

\subsection{Tuning to the Critical Point}

To start, let us look at how the lowest mass eigenvalues depend on the coupling $\lambar$. To do so, we truncate our conformal basis to some fixed $\Dmax$, keeping all states below this threshold, then diagonalize the lightcone Hamiltonian for various values of $\lambar$. Note that, unlike for the free field theory results in section~\ref{sec:SanityChecks}, here we include \emph{all} basis states with $\De \leq \Dmax$, regardless of particle number. Because each insertion of $\p_-\phi$ in a primary operator increases the scaling dimension by 1, this means we include states with up to $n=\Dmax$ particles.

Because we are only deforming our CFT by the even operators $\phi^2$ and $\phi^4$, the resulting spectrum can be divided into two independent sectors, depending on whether the eigenstates are odd or even under the $\mathbb{Z}_2$ transformation $\phi \ra -\phi$. In the following discussion, we identify the eigenvalues in these two sectors with the notation $\mu_{i,\odd/\even}^2$, where the label $i=1,2,\dots$ indicates the magnitude of the eigenvalue, with $i=1$ corresponding to the lowest eigenvalue in the respective $\mathbb{Z}_2$ sector.

\begin{figure}[t!]
\begin{center}
\includegraphics[width=0.7\textwidth]{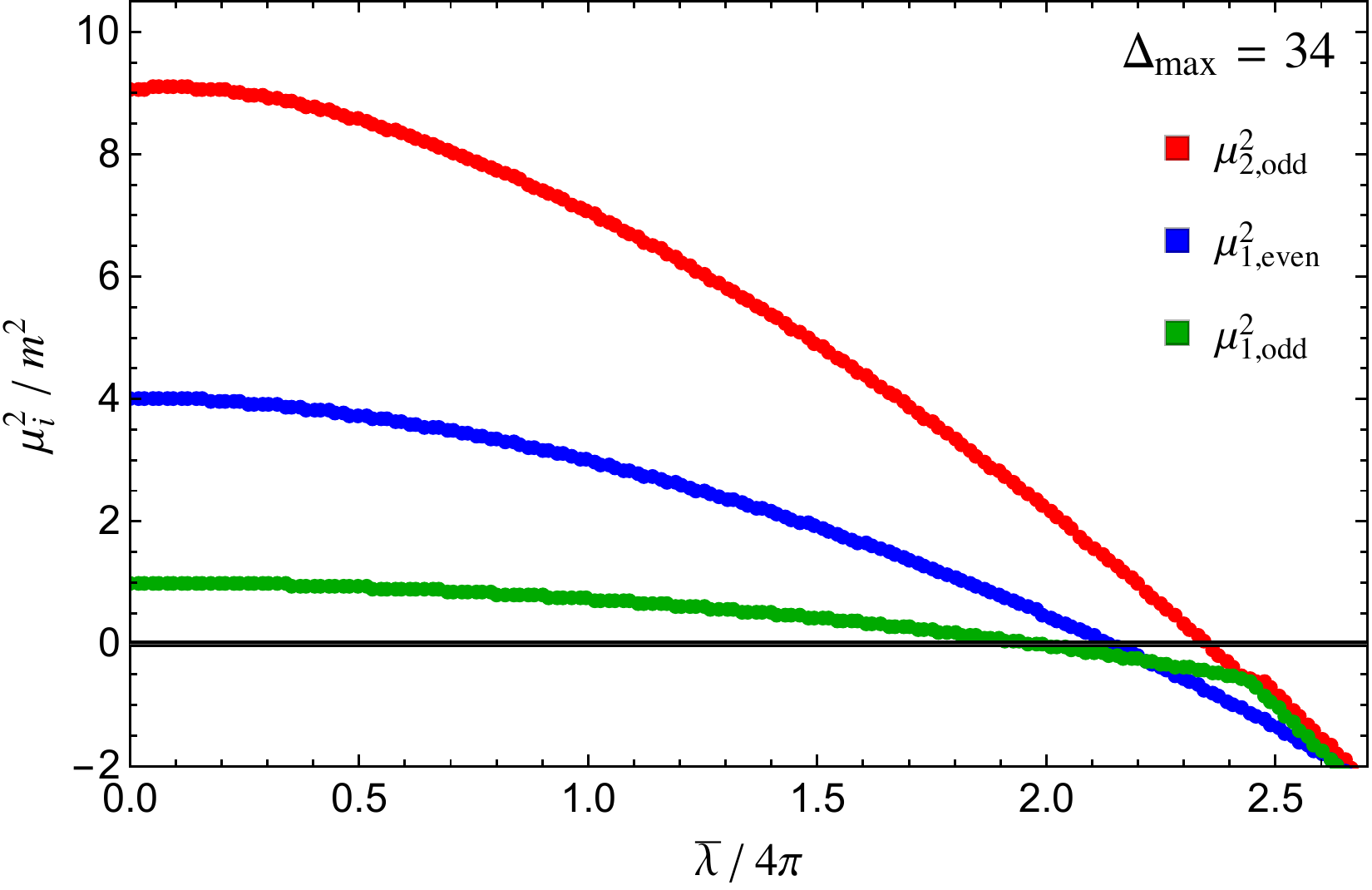}
\caption{The two lowest mass eigenvalues in the odd sector and the lowest eigenvalue in the even sector as a function of $\lambar$ for $\Dmax=34$ (12,310 basis states).} 
\label{fig:LowestEvalsDmax34} 
\end{center}
\end{figure}

Figure~\ref{fig:LowestEvalsDmax34} shows the lowest mass eigenvalues $\mu_{1,\odd}^2$, $\mu_{1,\even}^2$, and $\mu_{2,\odd}^2$ as functions of $\lambar$ for $\Dmax=34$, which corresponds to a basis of 12,310 states (the maximum truncation level we consider in this work). As we can see, at $\lambar=0$ these eigenvalues correspond to the 1-, 2-, and 3-particle thresholds, respectively. As the coupling $\lambar$ increases, all three of these eigenvalues begin to decrease, eventually reaching zero.\footnote{In particular, the mass eigenvalues cross zero and become negative. This is a signature of spontaneous symmetry-breaking in lightcone quantization~\cite{Rozowsky:2000gy}. In this work, we focus exclusively on the symmetry-preserving side of the critical point, leaving an analysis of the symmetry-broken phase for future work.}

Notice in figure~\ref{fig:LowestEvalsDmax34} that the eigenvalues go to zero at \emph{distinct} values of $\lambar$. This is clearly incorrect, as we expect the mass gap and the spacing between eigenvalues to all vanish at the same critical coupling. The discrepancy is due to truncation error, that is, a consequence of restricting our basis to finite $\Dmax$. We expect (and demonstrate below) that the discrepancy disappears in the limit $\Dmax\rightarrow\infty$. 

Even at finite $\Dmax$, though, our truncated data places a preliminary bound on the critical coupling, $\lambar_*$. Hamiltonian truncation is a type of variational method, which means that at any $\lambar$ the lowest eigenvalue ($\mu^2_{1,\odd}$) always places an \emph{upper bound} on the true mass gap. This in turn means that, for any finite $\Dmax$, the lowest eigenvalue reaches zero at a coupling strictly above the actual critical coupling. We can thus use the $\Dmax=34$ data to obtain the conservative bound
\be
\boxed{\fr{\lambar_*}{4\pi} \leq 1.98.}
\label{eq:bound}
\ee

To obtain the correct value for $\lambar_*$, we would like to extrapolate in $\Dmax$. To do this, we need to determine how the spectrum varies with $\Dmax$. At fixed $\lambar$, we find that the dependence of the lowest eigenvalues on $\Dmax$ is well modeled by 
\be
\mu_i^2(\Dmax) = A + \frac{B}{\De^n_{\max}},
\label{eq:fit}
\ee
where the parameters $A$, $B$, and $n$ are $\lambar$-dependent. In particular, the exponent $n$ tells us how quickly the truncation result for $\mu_i^2$ converges with $\Dmax$. We find experimentally that $n$ decreases monotonically with increasing $\lambar$, starting with $n\approx2$ at weak coupling and reaching $n\approx1$ near the critical point. This behavior for $n$ can be understood as a consequence of the Hamiltonian matrix elements' dependence on $\Dmax$, as we discuss in appendix~\ref{app:Decoupling}. By fixing $\lambar$ and varying $\Dmax$, we find the best fit for each mass eigenvalue $\mu_i^2$. The resulting parameter $A$ provides the extrapolated value of $\mu_i^2$ for that particular $\lambar$ in the limit $\Dmax\rightarrow\infty$.

\begin{figure}[t!]
\begin{center}
\includegraphics[width=\textwidth]{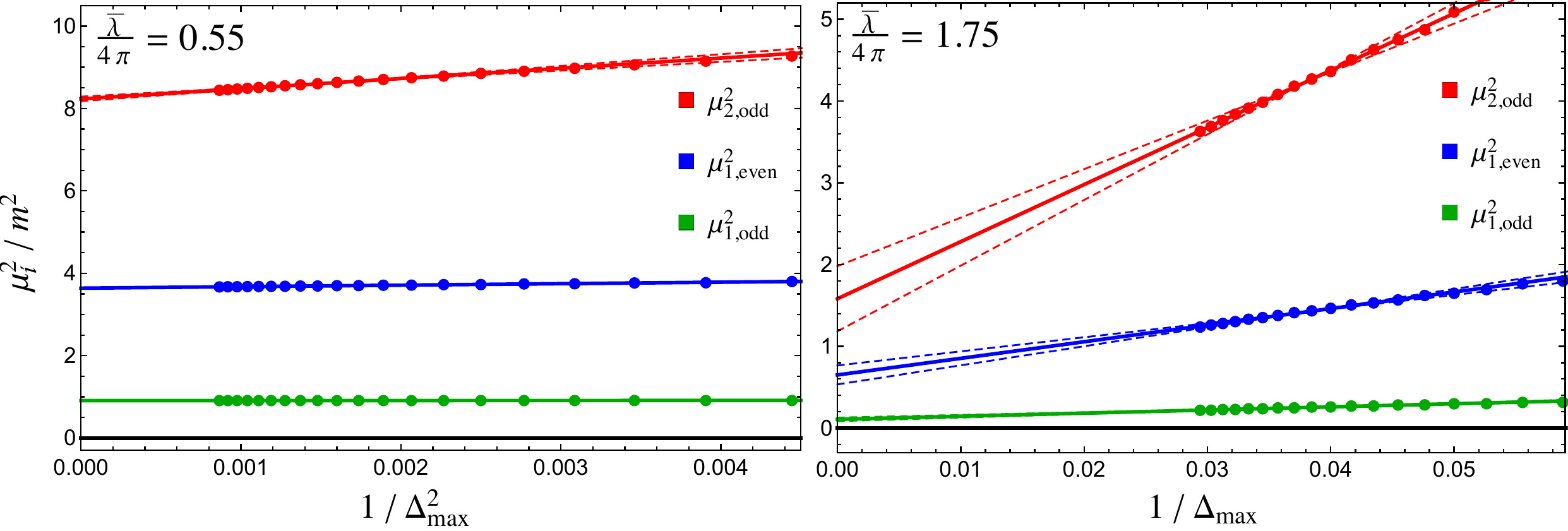}
\caption{Two examples of the dependence of $\mu_{1,\odd}^2$ (green), $\mu_{1,\even}^2$ (blue), and $\mu_{2,\odd}^2$ (red) on $\Dmax$, at fixed $\fr{\lambar}{4\pi}=0.55$ (left) and $\fr{\lambar}{4\pi}=1.75$ (right). The solid lines show the best fit for each $\mu_i^2(\Dmax)$ to the functional form in eq.~\eqref{eq:fit}, with the resulting powers $n=2.0$ (left) and $n=1.0$ (right). The $y$-intercept for each fit provides the extrapolated value of $\mu_i^2$ for $\Dmax\rightarrow\infty$, and the error is estimated by varying the slope by $15\%$ about the mean of the data points.} 
\label{fig:ExtrapolationExamples} 
\end{center}
\end{figure}

Figure~\ref{fig:ExtrapolationExamples} shows two examples of this procedure, one at $\fr{\lambar}{4\pi} = 0.55$ and the other at $\fr{\lambar}{4\pi} = 1.75$. The data points show the resulting values for $\mu_{1,\odd}^2$, $\mu_{1,\even}^2$, and $\mu_{2,\odd}^2$ at different $\Dmax$. The solid lines show the best fit for each $\mu_i^2(\Dmax)$, and the resulting $y$-intercept provides the extrapolated value as $\Dmax \ra \infty$. For the first example, which is clearly far from the critical point, we find that the corrections at finite $\Dmax$ fall as $1/\Dmax^n$ with $n=2.0$. The second example is much closer to criticality, and the results thus converge more slowly, with $n=1.0$.

The deviations of the data points from the best-fit curve are highly correlated, since increasing $\Dmax$ does not actually change any of the Hamiltonian matrix elements and instead just adds new ones. This correlation between data points makes it more difficult to determine the uncertainty in the extrapolated values for $\mu_i$, and standard estimates which ignore the correlation will typically underestimate the error in the resulting extrapolation. Rather than perform a detailed analysis of the uncertainty, we provide a simple estimate by varying the slope of the best fit line by $15\%$ about the mean of the data points, which corresponds to the dashed lines in figure~\ref{fig:ExtrapolationExamples}.

\begin{figure}[t!]
\begin{center}
\includegraphics[width=0.7\textwidth]{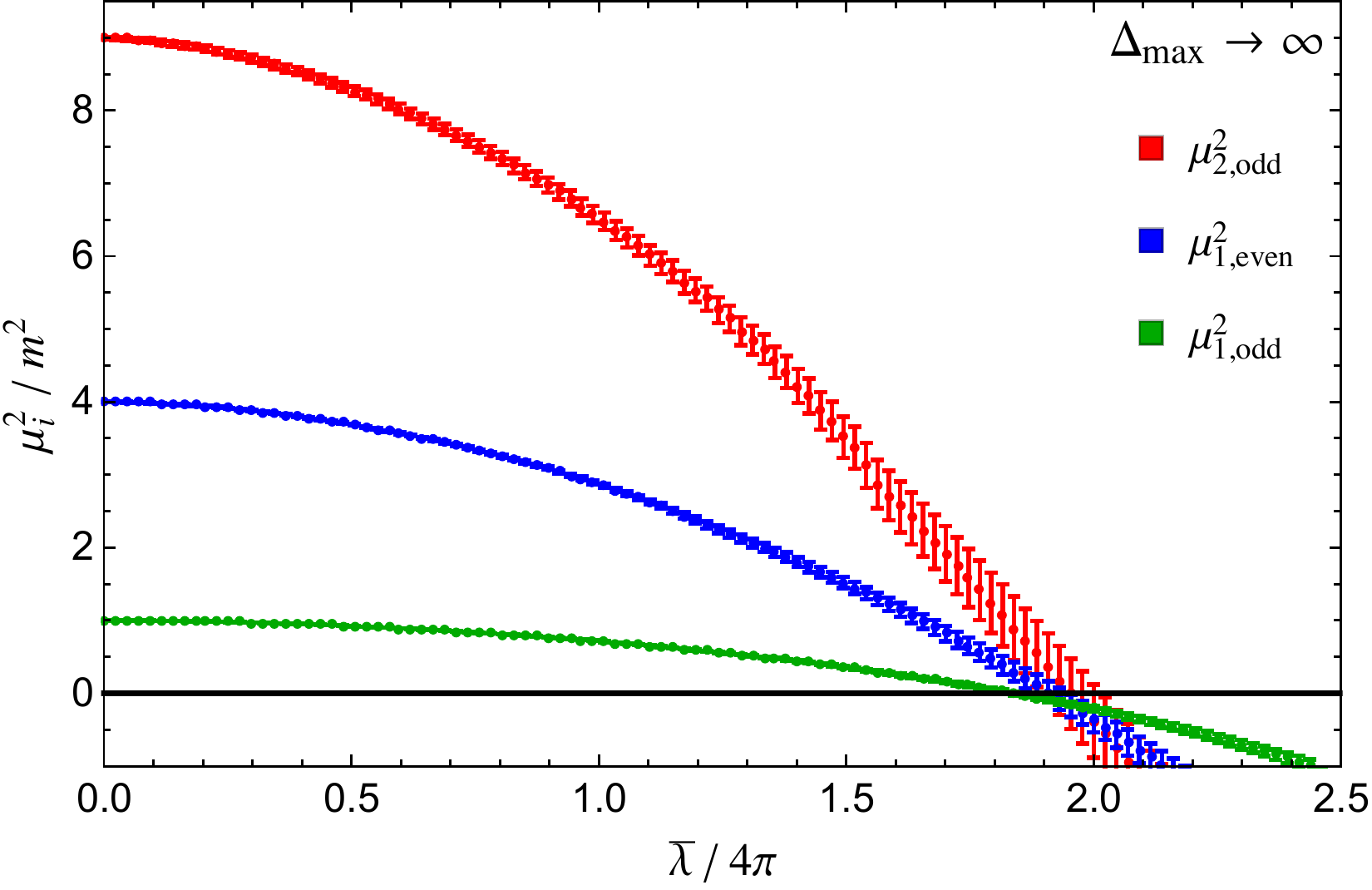}
\caption{The two lowest mass eigenvalues in the odd sector and the lowest eigenvalue in the even sector as a function of $\lb$ in the extrapolated limit $\Dmax\rightarrow\infty$.} 
\label{fig:LowestEvalsExtrapolate} 
\end{center}
\end{figure}

Carrying out this procedure for each $\lambar$, we are able to construct the $\Dmax\rightarrow\infty$ extrapolation for the lowest eigenvalues, shown in figure~\ref{fig:LowestEvalsExtrapolate}. This plot is the analogue of figure~\ref{fig:LowestEvalsDmax34}, showing the extrapolated values for $\mu_{1,\odd}^2$, $\mu_{1,\even}^2$, and $\mu_{2,\odd}^2$ as a function of $\lambar$. We see that, unlike at finite $\Dmax$, all three eigenvalues reach zero at the same $\lambar$, to within the error bars.

\begin{figure}[t!]
\begin{center}
\includegraphics[width=0.7\textwidth]{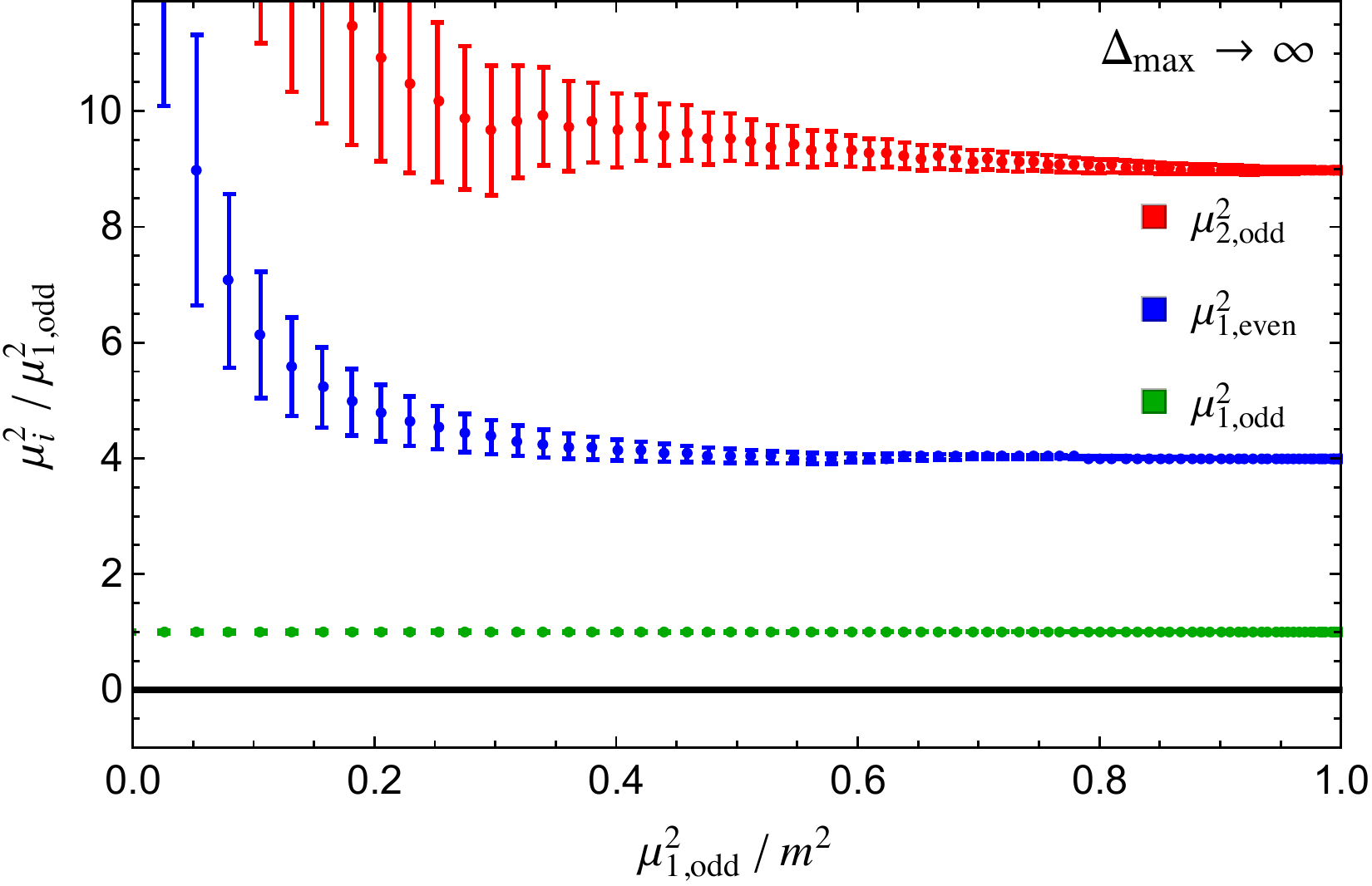}
\caption{The ratio of two lowest mass eigenvalues in the odd sector and the lowest eigenvalue in the even sector to the mass gap as a function of $\mu^2_{1,\odd}$ in the extrapolated limit $\Dmax\rightarrow\infty$.} 
\label{fig:MassRatios} 
\end{center}
\end{figure}

We can now use these extrapolated eigenvalues to determine the critical coupling. Our best estimate clearly comes from the lowest eigenvalue, $\mu_{1,\odd}^2$, which has the least uncertainty in its extrapolation. By measuring the point at which this eigenvalue reaches zero, we obtain the prediction
\be
\boxed{\fr{\lambar_*}{4\pi} = 1.84 \pm 0.03.}
\ee

As another simple check of this extrapolation, figure~\ref{fig:MassRatios} shows the extrapolated ratios of the eigenvalues $\mu_{1,\even}^2$ and $\mu_{2,\odd}^2$ to the mass gap $\mu_{1,\odd}^2$, as a function of the gap. We see that, although the eigenvalues themselves change significantly, their ratios appear to remain fixed at the free field values of $1$, $4$, and $9$, corresponding to the one-, two-, and three-particle thresholds. This matches our expectation that there should be no bound states in $\phi^4$ theory.

However, the ratios begin to deviate from the expected values as we near the critical point, indicating that the one-particle state still reaches zero before the two- and three-particle thresholds. This deviation is due to the fact that we have extrapolated these ratios from results with finite $\Dmax$, which limits our IR resolution. These ratios thus provide a useful indicator of the approximate scale of our IR cutoff.

%%%%%%%%%%%%%%%%%%%%%%%%%%%%%%%%%%%%%%%%%%%%%%%%%%%%%%%%%%%%%%%%%%%%%%%%%%%%%

\subsection{Comparison with Prior Work}
\label{subsec:compare}

The critical coupling of 2D $\phi^4$ theory has been studied previously using a variety of computational methods in both lightcone~\cite{Harindranath:1987db,Harindranath:1988zt,Burkardt:2016ffk} and equal-time quantization~\cite{Lee:2000ac,Sugihara:2004qr,Schaich:2009jk,Milsted:2013rxa,Bosetti:2015lsa,Rychkov:2014eea}. As we briefly summarize below, the value of the critical coupling is dependent on the choice of quantization scheme, such that mapping between the lightcone and equal-time values is rather difficult. We do not attempt a comparison with equal-time results in this work, since it is somewhat tangential to our main goal, and instead focus on comparing our result for $\lambar_*$ with values from other lightcone methods. While $\lambar_*$ is certainly an important intermediate result of this work, ultimately we are interested in computing physical observables like correlation functions which, unlike the critical coupling, are independent of quantization scheme.

The first study of the critical coupling in lightcone quantization appeared in \cite{Harindranath:1987db,Harindranath:1988zt}. This work used the method of discretized lightcone quantization (DLCQ) \cite{Pauli:1985pv,Pauli:1985ps,Brodsky:1997de}, which is a Hamiltonian truncation method where the underlying QFT Hilbert space is discretized by compactifying the ``spatial'' lightcone direction $x^-$. More recently, the critical coupling was studied in \cite{Burkardt:2016ffk} using a Hamiltonian truncation method with a basis of symmetric polynomials in momentum space. These results for the critical coupling, along with ours, are summarized below:
{\setlength{\extrarowheight}{2pt}%
\begin{center}
\begin{tabular}{ |l|l| } 
 \hline
 Lightcone Method &\, $\bar{\lambda}_* / (4\pi)$ \\ 
 \hline
 DLCQ~\cite{Harindranath:1988zt} &\, $2.6$ \\ 
 Symmetric polynomials~\cite{Burkardt:2016ffk} &\, $2.1 \pm .05$ \\
 Conformal truncation (this work) \hspace{5mm} &\, $1.84 \pm .03$ \\
 \hline
\end{tabular}
\end{center}

Our extrapolated value for the critical coupling is somewhat lower than the values obtained in both \cite{Harindranath:1988zt} and \cite{Burkardt:2016ffk}. There is also some tension between these previous results and our data even before we perform any extrapolation in $\Dmax$. Recall from our discussion above that conformal truncation is a variational method, so that our $\Dmax = 34$ data places an explicit upper bound on the value of the critical coupling, $\fr{\lambar_*}{4\pi} \leq 1.98$. The values reported in \cite{Harindranath:1988zt,Burkardt:2016ffk} are centered above this bound.

Ref.~\cite{Harindranath:1988zt} is an older work and does not report error bars, so it is difficult to ascertain the precision of this result for comparison. As for~\cite{Burkardt:2016ffk}, their basis of symmetric polynomials has a one-to-one map to the basis states we use in this work (see appendix~\ref{app:Basis}). For this particular theory, our methods are thus completely equivalent in practice, although there are minor technical differences in actual implementation. The maximum basis size considered in~\cite{Burkardt:2016ffk} consists of 226 total states and corresponds to a subset of our $\Dmax=18$ basis. For our results, we have constructed the basis up to $\Dmax=34$, which consists of 12,310 states. It is thus possible that the uncertainty in these previous results is somewhat larger than initially estimated, which would allow for compatibility with our higher $\Dmax$ results.     

Comparison with equal-time results is more subtle, because the value of the critical coupling is quantization scheme dependent. The difference between the two schemes can be seen most easily at the level of Feynman diagrams: there exist mass-renormalization diagrams due to the coupling $\lambar$ that appear in equal-time quantization but vanish in lightcone quantization~\cite{Burkardt:1992sz}. A given value of the bare coupling $\lambar$ thus clearly leads to different physical masses in the two quantization schemes. In principle, it should be possible to resum the missing diagrams in order to convert between lightcone and equal-time results, and ref.~\cite{Burkardt:2016ffk} proposes such a method. This prescription, however, is inherently non-perturbative due to the need to account for an infinite class of diagrams. While outside the scope of this current work, it would be very interesting and instructive to perform a careful matching between lightcone and equal-time data and to compare our results to those reported in~\cite{Lee:2000ac,Sugihara:2004qr,Schaich:2009jk,Milsted:2013rxa,Bosetti:2015lsa,Rychkov:2014eea}.

%%%%%%%%%%%%%%%%%%%%%%%%%%%%%%%%%%%%%%%%%%%%%%%%%%%%%%%%%%%%%%%%%%%%%%%%%%%%%
%%%%%%%%%%%%%%%%%%%%%%%%%%%%%%%%%%%%%%%%%%%%%%%%%%%%%%%%%%%%%%%%%%%%%%%%%%%%%

\section{Ising Model Near Critical Temperature}
\label{sec:Ising}

Now that we have confirmed the existence of a critical point and determined the corresponding critical coupling, $\lambar_*$, we can turn to the main focus of this work: computing dynamical observables, namely spectral densities, in the vicinity of the fixed point. This IR fixed point is described by the 2D Ising model near the critical temperature $T_c$,
\be
\Lcal = \half \p^\mu\phi \p_\mu\phi - \half m^2 \phi^2 - \fr{1}{4!} \lambda \phi^4 \hspace{3mm}\Rightarrow\hspace{3mm} \Lcal_{\textrm{Ising}} - m_{\gap} \ep,
\ee
where the arrow denotes RG flow to the IR. Here $m_{\gap}\ra 0$ as $\lambar \ra \lambar_*$, and the deformation by $\ep$ is equivalent to moving the Ising model away from the critical temperature $T_c$, with $m_\gap \sim |T-T_c|$. This IR theory is famously integrable, such that one can compute its spectral densities analytically. In this section, we use conformal truncation to compute spectral densities in $\phi^4$ theory for any $\lambar$, then verify that near $\lambar_*$ they match the known analytic results for the Ising model, allowing us to test our method in a strongly-coupled example.

Recall that we compute spectral densities by first truncating the basis to some $\Dmax$ and then numerically diagonalizing the resulting lightcone Hamiltonian matrix to obtain the approximate mass eigenstates $|\mu_i\>$. The integrated spectral density of any operator is then given by eq.~(\ref{eq:I}). 

Specifically, we compute and study the spectral densities of the stress-energy tensor $T_{\mu\nu}$ and the scalar operators $\phi^n$. These operators are all initially defined in the UV. For the stress tensor, we can study the spectral densities of individual components. A particularly interesting component is $T_{+-}$, which in 2D is proportional to the trace,
\benn
T^\mu_{\phantom{\mu}\mu} = 2 T_{+-}.
\eenn
The theoretical prediction for this particular component is that near criticality
\be
T_{+-} \Rightarrow m_{\gap} \ep.
\ee
Note that this vanishes at the critical coupling, since $m_{\gap} \ra 0$. By computing the spectral density of $T_{+-}$, we are thus able to explicitly check whether the stress tensor is traceless at $\lambar_*$, which determines whether the critical point corresponds to a CFT. The ability to study the RG flow of the stress tensor is a particularly useful feature of conformal truncation, as other non-perturbative methods typically break translation invariance, making it difficult to reproduce the stress tensor.

For the $\phi^n$ operators, the expectation is that near criticality their IR description will be in terms of the leading operators in the Ising model, namely, $\s$ (the lowest $\mathbb{Z}_2$-odd operator) and $\ep$ (the lowest $\mathbb{Z}_2$-even operator). Near the critical point $\lambar_*$, we thus expect the \emph{universal} behavior
\be
\phi,\phi^3,\phi^5,\ldots \Rightarrow \s, \qquad \phi^2,\phi^4,\phi^6,\ldots \Rightarrow \ep.
\ee
In other words, we expect that near $\lambar_*$ the $\mu \ra 0$ behavior of the spectral densities $\rho_{\phi^n}$ will approach the known expressions for $\rho_\s$ or $\rho_\ep$, depending on parity.

While not technically an independent degree of freedom (due to the Ward identity), the component $T_{--}$ of the stress tensor is also a useful observable. Its integrated spectral density is equivalent to the Zamolodchikov $C$-function, which measures the change in central charge between the UV and IR fixed points and is an intrinsic feature of the intermediate RG flow. Using conformal truncation, we can compute the $C$-function at any coupling $\lambar$. Compared to $T_{+-}$ and $\phi^n$, however, it is more difficult to extract the Ising model behavior near criticality from $T_{--}$ due to its sensitivity to corrections from UV physics, as we discuss.

%%%%%%%%%%%%%%%%%%%%%%%%%%%%%%%%%%%%%%%%%%%%%%%%%%%%%%%%%%%%%%%%%%%%%%%%%%%%%

\subsection{Trace of the Stress-Energy Tensor}
\label{subsec:stress}

To begin, let us consider the trace of the stress-energy tensor. In 2D, the trace is proportional to the component $T_{+-}$, which for $\phi^4$ theory takes the form (see section~\ref{sec:SanityChecks})
\be
T_{+-} = \half m^2 \phi^2 + \fr{1}{4!} \lambda \phi^4 + \fr{1}{16\pi} \lambda \phi^2.
\ee
Near the critical coupling, we expect $T_{+-}$ to match onto the 2D Ising prediction in the IR. Exact predictions for the Ising model at $T\neq T_c$ are possible, because the theory is integrable and can be described in terms of a free fermion with mass $m_\gap$. The Ising spectral density for $T_{+-}$ can be computed analytically from its decomposition into Fock space states with two fermions \cite{Cappelli:1989yu},
\be
\rho_{T_{+-}}(\mu) = \fr{m_\gap^2}{2!} \int \fr{d\theta_1 d\theta_2}{(4\pi)^2} \, (2\pi) \de^2(P-p_1-p_2) \sinh^2 \fr{\theta_{12}}{2} = \fr{m_\gap^2}{16\pi} \sqrt{1 - \fr{4m_\gap^2}{\mu^2}} \quad \, \, (T > T_c),
\ee
where $\theta$ is the rapidity of an individual fermion with $p_{\pm} = m_{\gap} e^{\pm \theta}$, and $\theta_{ij} \equiv \theta_i - \theta_j$.  Near the critical coupling $\lambar_*$, we therefore expect the spectral density of $T_{+-}$ to flow to this Ising model prediction in the IR. In particular, recall that $m_{\gap} \ra 0$ as $\lambar \ra \lambar_*$, so the spectral density should vanish as we approach criticality, as expected for an IR CFT. 

\begin{figure}[t!]
\begin{center}
\includegraphics[width=\textwidth]{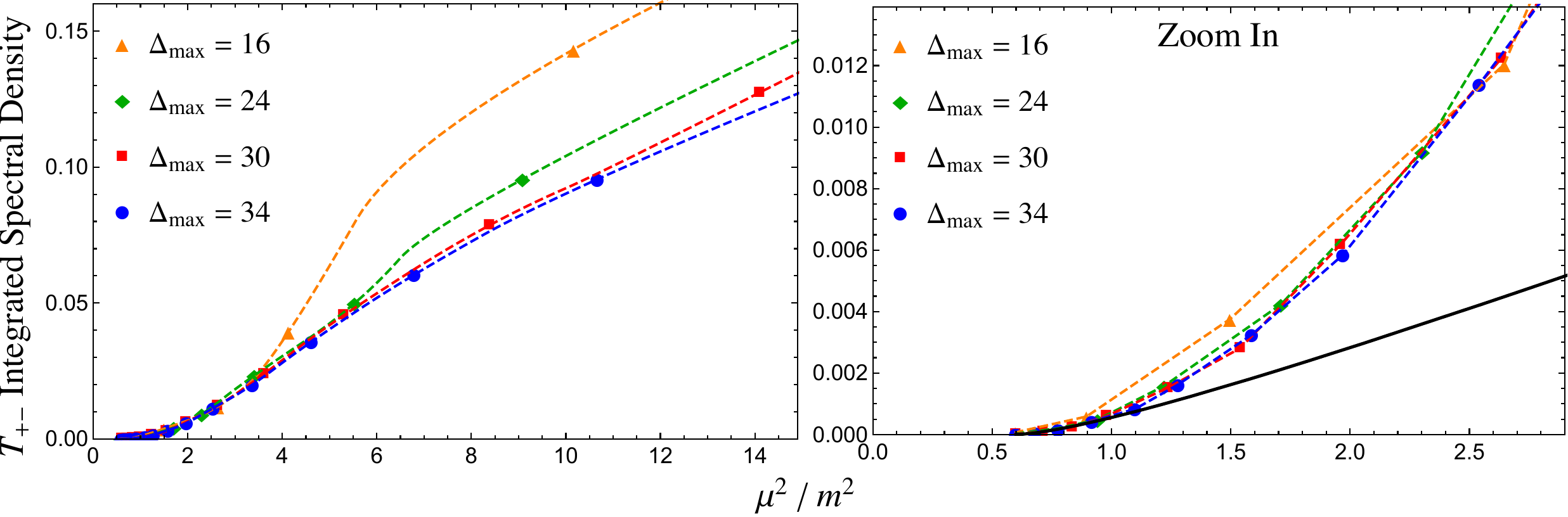}
\caption{Integrated spectral density for $T_{+-}$ at different values of $\Dmax$. The $\Dmax=34$ results (blue dots) are at $\fr{\lambar}{4\pi} = 1.96$, and the couplings for the remaining results have been chosen such that the mass gap remains fixed. The points are the actual contributions of individual eigenstates to the spectral density, while the dashed lines are interpolations. The right plot is simply a zoomed-in version of the left one, and compares the conformal truncation results to the theoretical IR prediction for the Ising model (black curve).} 
\label{fig:TraceConvergence} 
\end{center}
\end{figure}

Before comparing the conformal truncation integrated spectral densities with the predictions from the Ising model, we can study their behavior as a function of $\Dmax$ to determine how quickly the results converge. When comparing results with different values of $\Dmax$, we have a choice as to which parameter to hold fixed. One obvious choice is to fix the coupling $\lambar$ (as we did in the extrapolations in section~\ref{sec:Coupling}), in which case the IR scale $m_\gap$ will vary as we increase $\Dmax$. Alternatively, we can hold $m_\gap$ fixed and vary $\lambar$. Because we are specifically interested in studying IR dynamics, we choose the latter option, keeping $m_\gap$ fixed in order to study the convergence of our results relative to this physical IR scale.

Figure~\ref{fig:TraceConvergence} shows our truncation results for the integrated spectral density of $T_{+-}$ at four different values of $\Dmax$. The results with the highest truncation level, $\Dmax=34$, are at $\fr{\lambar}{4\pi} = 1.96$. For the results with lower $\Dmax$, the couplings have thus been chosen to ensure that in each case the mass gap matches that of the $\Dmax=34$ spectrum.

As we can see, the $\Dmax=34$ results appear to have converged across a wide range of mass scales, suggesting that these results are successfully computing the true spectral densities. Moreover, we see that conformal truncation appears to reconstruct the spectral densities from the IR up, such that even $\Dmax=16$ is an accurate approximation to the low-energy dynamics. This behavior appears to confirm our intuition that states with low conformal Casimir in the UV provide the dominant contribution to low-mass states, even at strong coupling.

In the right plot of figure~\ref{fig:TraceConvergence}, we compare our truncation results to the theoretical prediction for the Ising model (black curve). This analytic expression only has one unknown parameter, $m_\gap$, which is fixed by setting the lowest eigenvalue $\mu_{1,\even}^2 = 4 m_\gap^2$. In the IR, the conformal truncation results clearly match both the scaling and overall coefficient of the Ising model prediction. 

It is worth emphasizing that the correspondence between $\phi^4$ theory and Ising model spectral densities should only hold in the deep IR. At higher energy scales $\mu^2$, these theories are not equivalent and thus have distinct spectral densities, which is precisely what we observe in figure~\ref{fig:TraceConvergence}.

\begin{figure}[t!]
\begin{center}
\includegraphics[width=1\textwidth]{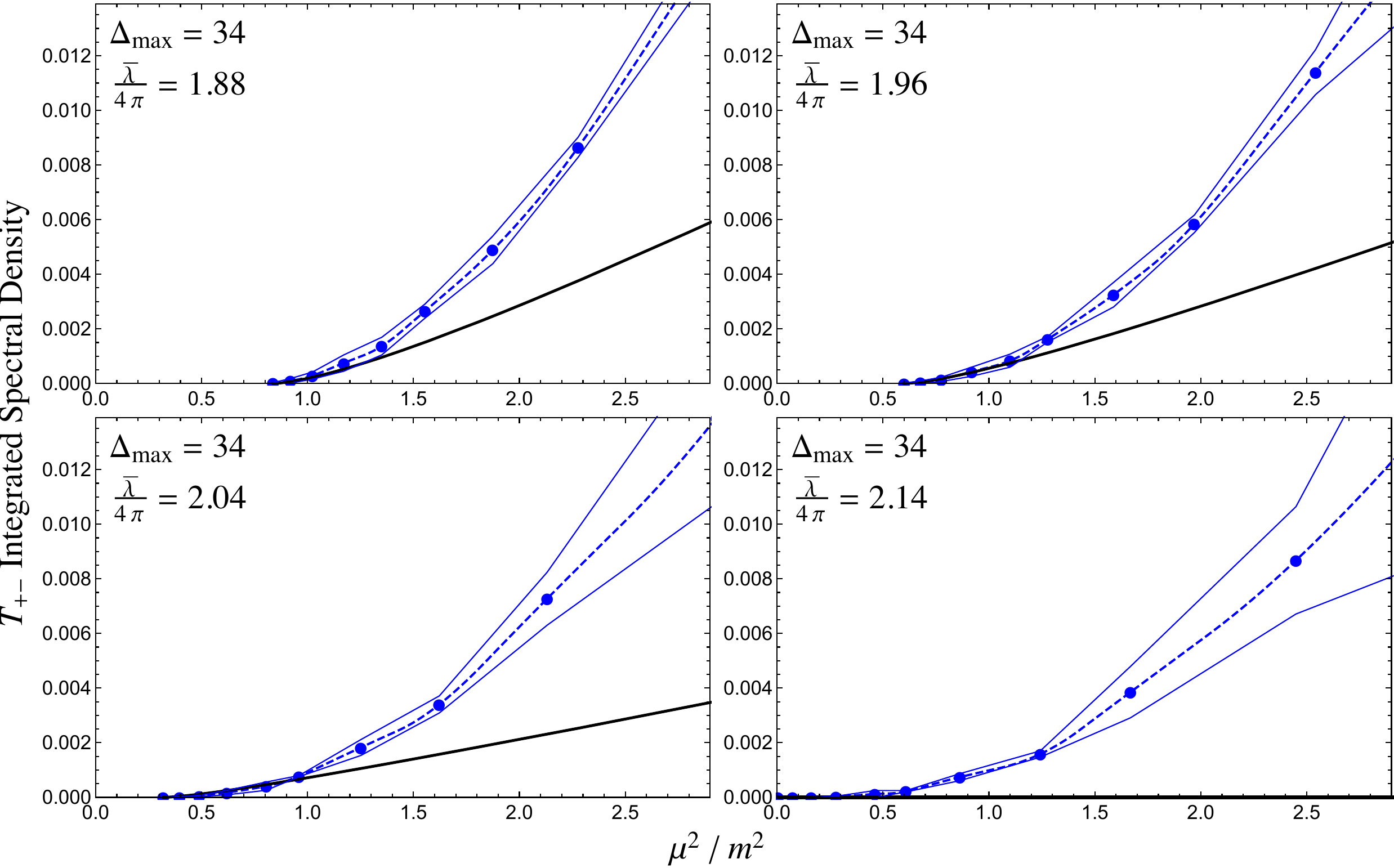}
\caption{Integrated spectral densities for $T_{+-}$, for $\Dmax = 34$ and different values of $\lambar$, compared to the Ising model prediction (black curve). The thin blue lines indicate the magnitude of the difference between these results and those at $\Dmax=30$, providing a rough estimate of the convergence. For reference, the upper right plot corresponds to the same value of the coupling ($\fr{\lambar}{4\pi} = 1.96$) as figure~\ref{fig:TraceConvergence}.} 
\label{fig:TraceZoomIn} 
\end{center}
\end{figure}

Figure~\ref{fig:TraceZoomIn} shows the $\Dmax=34$ results for the $T_{+-}$ spectral density at multiple values of $\lambar$ near the critical point, again compared to the theoretical prediction from the Ising model. As a rough estimate of the convergence, we have included an envelope surrounding the truncation results whose width corresponds to the difference between these results and those at $\Dmax=30$. We see that the spectral density correctly reproduces the Ising model prediction in the IR over a range of couplings. Most importantly, the resulting IR density \emph{vanishes} as $m_\gap \ra 0$, clearly indicating that the critical theory is described by a CFT.

While this is not surprising, as we already know that the critical point of $\phi^4$ theory should be described by the 2D Ising model, this example demonstrates the utility of spectral densities in analyzing the low-energy behavior of strongly-coupled theories. For more general RG flows, where the IR description is unknown, seeing the trace of the stress tensor vanish in conjunction with the mass gap confirms that the UV theory flows to an IR CFT.

The spectral density of the stress tensor trace also clearly delineates which eigenstates correspond to the IR fixed point. As we can see in figure~\ref{fig:TraceZoomIn}, the spectral density is zero for roughly the first six points, indicating that these states comprise the IR sector described by the critical Ising model.

%%%%%%%%%%%%%%%%%%%%%%%%%%%%%%%%%%%%%%%%%%%%%%%%%%%%%%%%%%%%%%%%%%%%%%%%%%%%%

\subsection{Universality in $\phi^n$ Spectral Densities}

Next, we can turn to the scalar operators $\phi^n$. Near the critical coupling $\lambar_*$, we expect that in the IR these operators will all flow to the lowest dimension operators in the Ising model,
\be
\phi^{2n} \Rightarrow \ep + \cdots, \quad \phi^{2n-1} \Rightarrow \s + \cdots,
\ee
where the ellipses denote higher-dimensional operators. We thus expect \emph{universal} behavior in the associated spectral densities as $\mu\ra 0$,
\be
\rho_{\phi^{2n}}(\mu) \ra \rho_\ep(\mu), \quad \rho_{\phi^{2n-1}}(\mu) \ra \rho_\s(\mu) \qquad (\mu \ra 0).
\ee

The theoretical prediction for the $\ep$ spectral density is identical to that of $T_{+-}$, but without the overall factor of $m_\gap^2$,
\be
\rho_\ep(\mu) = \fr{1}{16\pi} \sqrt{1 - \fr{4m_\gap^2}{\mu^2}} \qquad (T > T_c).
\ee
On the other hand, $\s$ has overlap with all Fock space states with odd numbers of fermions, leading to the more complicated spectral density \cite{Berg:1978sw,McCoy:1977er,Wu:1975mw},
\be
\rho_\s(\mu) = \sum_{n \textrm{ odd}} \fr{1}{n!} \int \prod_{k=1}^n \left(\fr{d\theta_k}{4\pi}\right) (2\pi) \de^2\Big(P-\sum_k p_k\Big) \, 2^{n-1} \prod_{i \leq j} \tanh^2 \fr{\theta_{ij}}{2} \qquad (T > T_c).
\label{eq:rhos}
\ee
However, the contribution of each $n$-fermion sector begins at $\mu = n m_\gap$, which means that in practice we only need to consider the contributions from the states with low fermion number to determine the IR behavior. Moreover, for the mass scales $\mu^2$ that we consider, the overwhelmingly dominant term is the single-fermion contribution, which is a delta function. Thus, the $\s$ integrated spectral density is simply a step function at $\mu^2=m_\gap^2$, with only sub-percent level corrections coming from higher fermion number contributions.

\begin{figure}[t!]
\begin{center}
\begin{tabular}{c}
\includegraphics[width=0.95\textwidth]{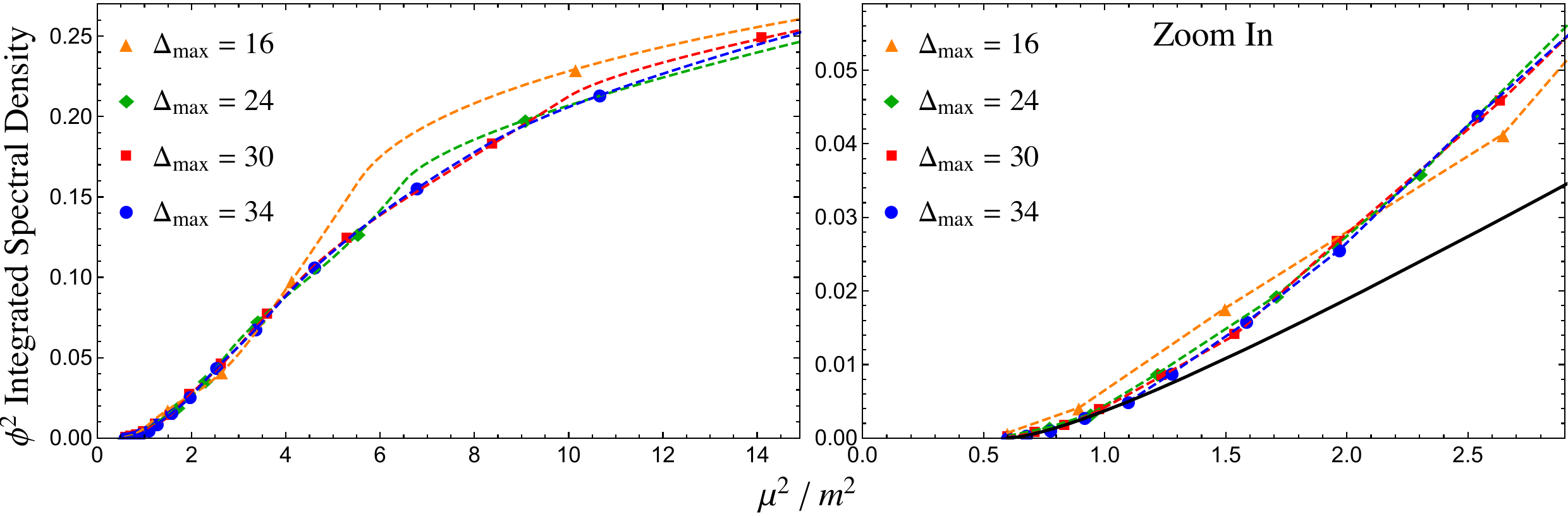} \\
\includegraphics[width=0.95\textwidth]{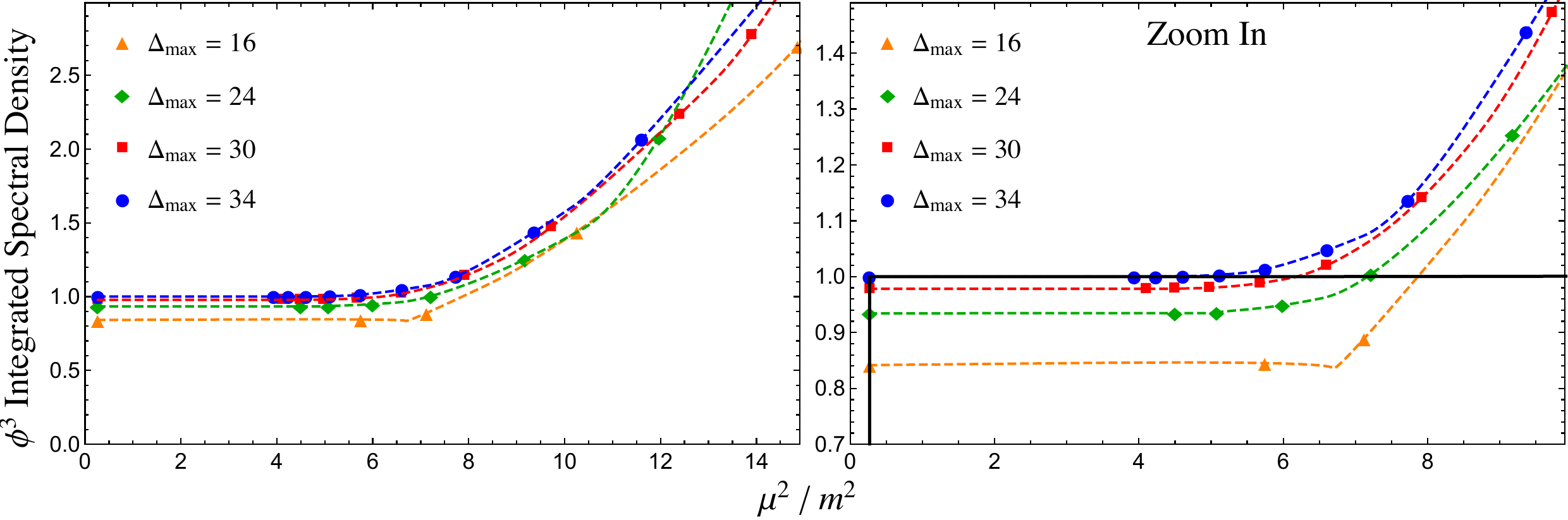}
\end{tabular}
\caption{Integrated spectral densities for $\phi^2$ (top) and $\phi^3$ (bottom) at different values of $\Dmax$. The $\Dmax=34$ results (blue dots) are at $\fr{\lambar}{4\pi} = 1.96$ (top) and $\fr{\lambar}{4\pi} = 1.69$ (bottom), and the couplings for the remaining results have been chosen such that the respective mass gaps remain fixed. The points are the actual contributions of individual eigenstates to the spectral density, while the dashed lines are interpolations. The right plots are simply a zoomed-in version of the left ones, and compare the conformal truncation results to the theoretical IR predictions for $\ep$ (top) and $\sigma$ (bottom) in the Ising model.} 
\label{fig:PhiNConvergence} 
\end{center}
\end{figure}

\begin{figure}[t!]
\begin{center}
\begin{tabular}{c c}
\includegraphics[width=0.475\textwidth]{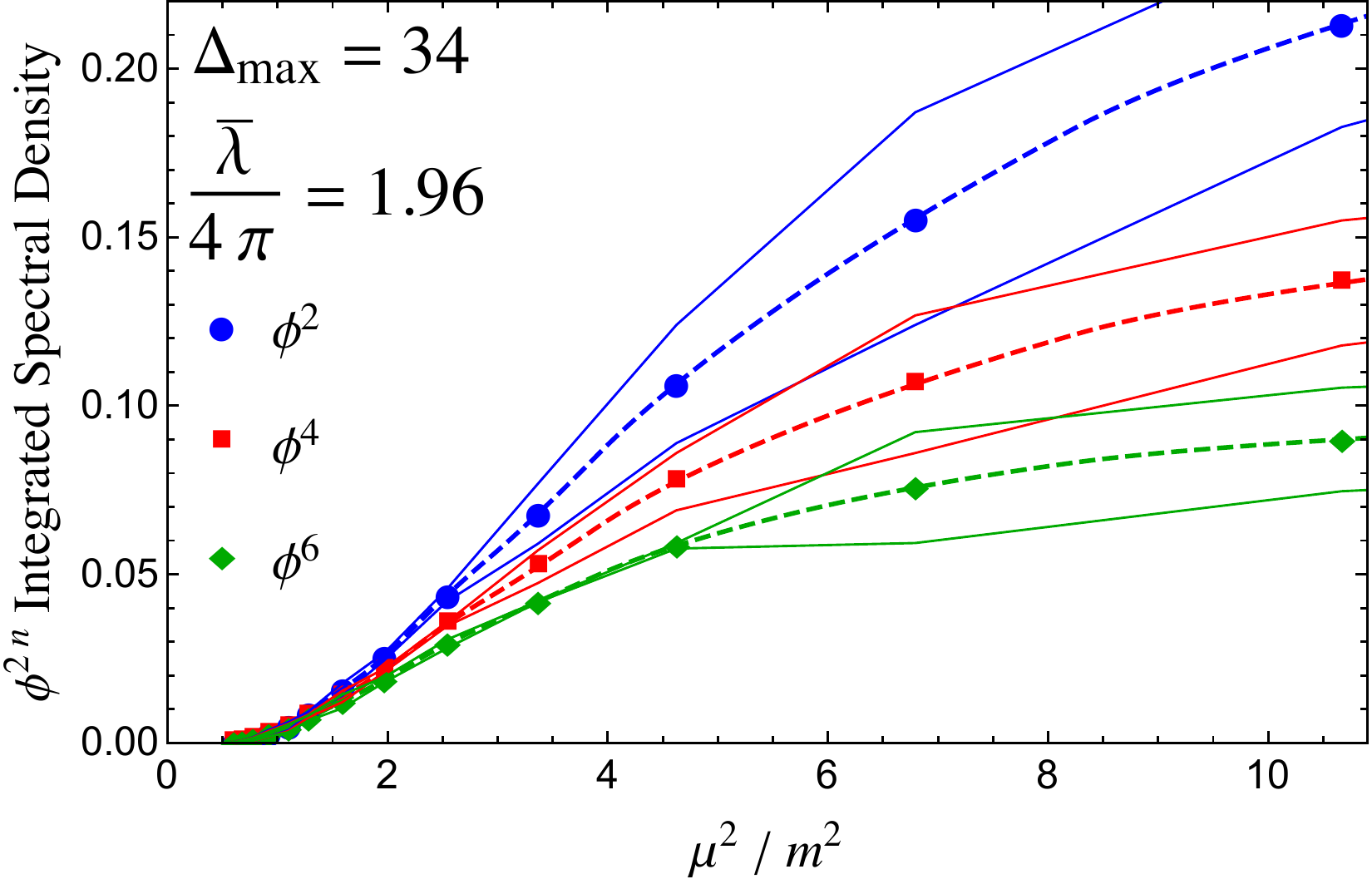} &
\includegraphics[width=0.475\textwidth]{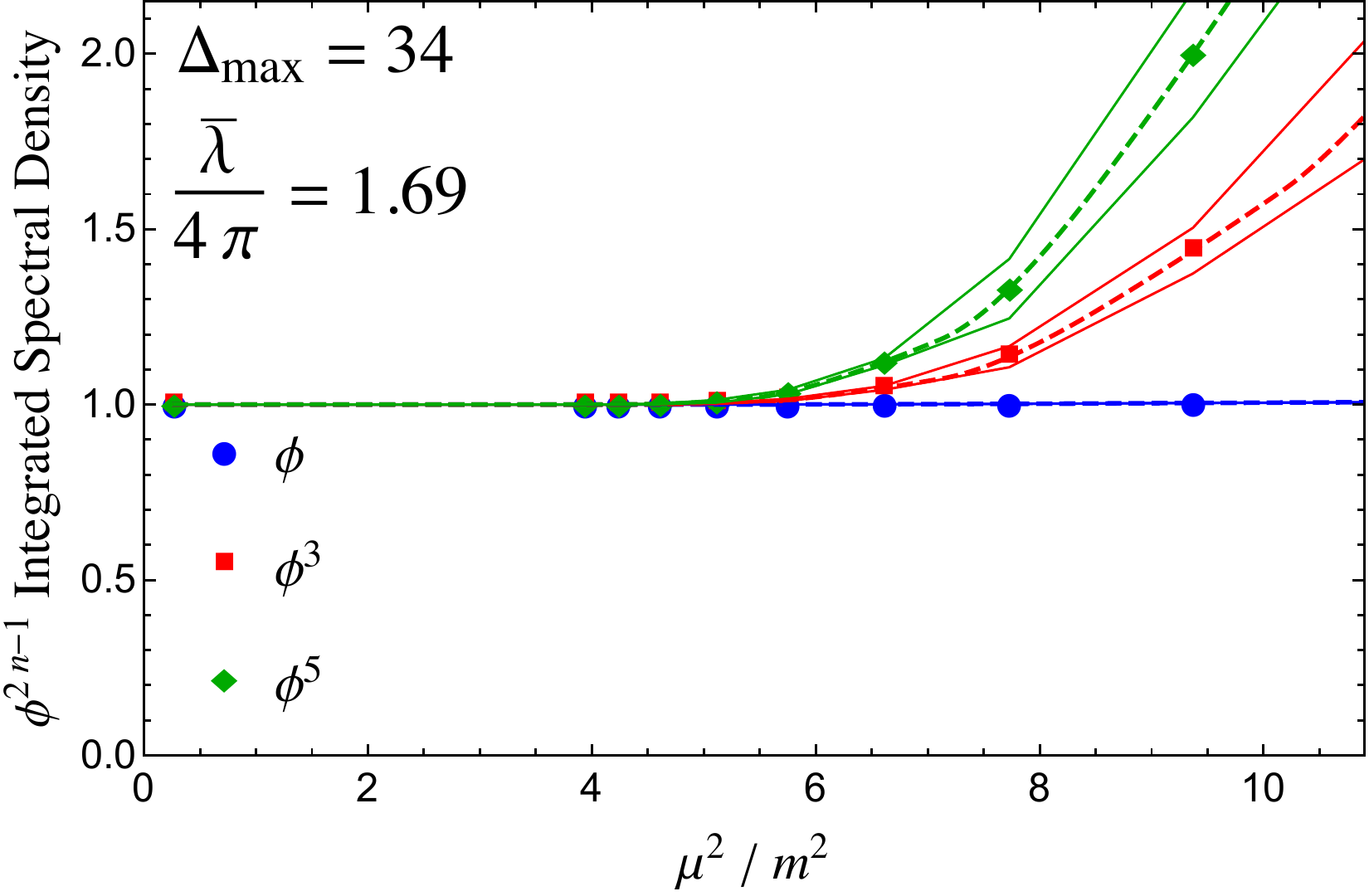}
\end{tabular}
\caption{Integrated spectral densities for $\phi^2$, $\phi^4$, and $\phi^6$ at $\fr{\lambar}{4\pi} = 1.96$ (left) and for $\phi$, $\phi^3$, and $\phi^5$ at $\fr{\lambar}{4\pi} = 1.69$ (right), both with $\Dmax = 34$. The spectral densities in each plot have been rescaled by an overall coefficient such that the first data points match. The thin lines indicate the magnitude of the difference between these results and those at $\Dmax=30$, providing a rough estimate of the convergence. In both plots, all three curves converge to the same universal behavior in the IR.} 
\label{fig:PhiNZoomOut}
\end{center}
\end{figure}

Just like with $T_{+-}$, we first study the rate of convergence by plotting the $\phi^n$ spectral densities at various $\Dmax$ with fixed $m_\gap$, as shown in figure~\ref{fig:PhiNConvergence}. These plots specifically show $\phi^2$ and $\phi^3$, with similar results for the other operators. For the highest truncation level, $\Dmax=34$, the coupling was fixed to $\fr{\lambar}{4\pi} = 1.96$ for $\phi^2$ and $\fr{\lambar}{4\pi} = 1.69$ for $\phi^3$.

We again find that the conformal truncation results converge rather quickly, especially in the IR. The rightmost plots compare the low-mass results to the theoretical predictions for $\ep$ and $\s$ (black curves). Note that these spectral densities are merely expected to be \emph{proportional} to those of $\ep$ and $\s$ in the IR, with an unknown $\lambar$-dependent overall coefficient for each $\phi^n$. These coefficients can be fixed by fitting the overall normalization of the $\phi^n$ spectral densities to the theoretical predictions. Because we only expect these operators to match the Ising predictions in the IR, we specifically fit the normalization to the lowest 5 data points. As we can see, both operators match their Ising model predictions at low energies. This is especially noticeable for the $\phi^3$ spectral density, which develops a large resonance corresponding to the one-fermion contribution to $\s$.

\begin{figure}[t!]
\begin{center}
\includegraphics[width=1\textwidth]{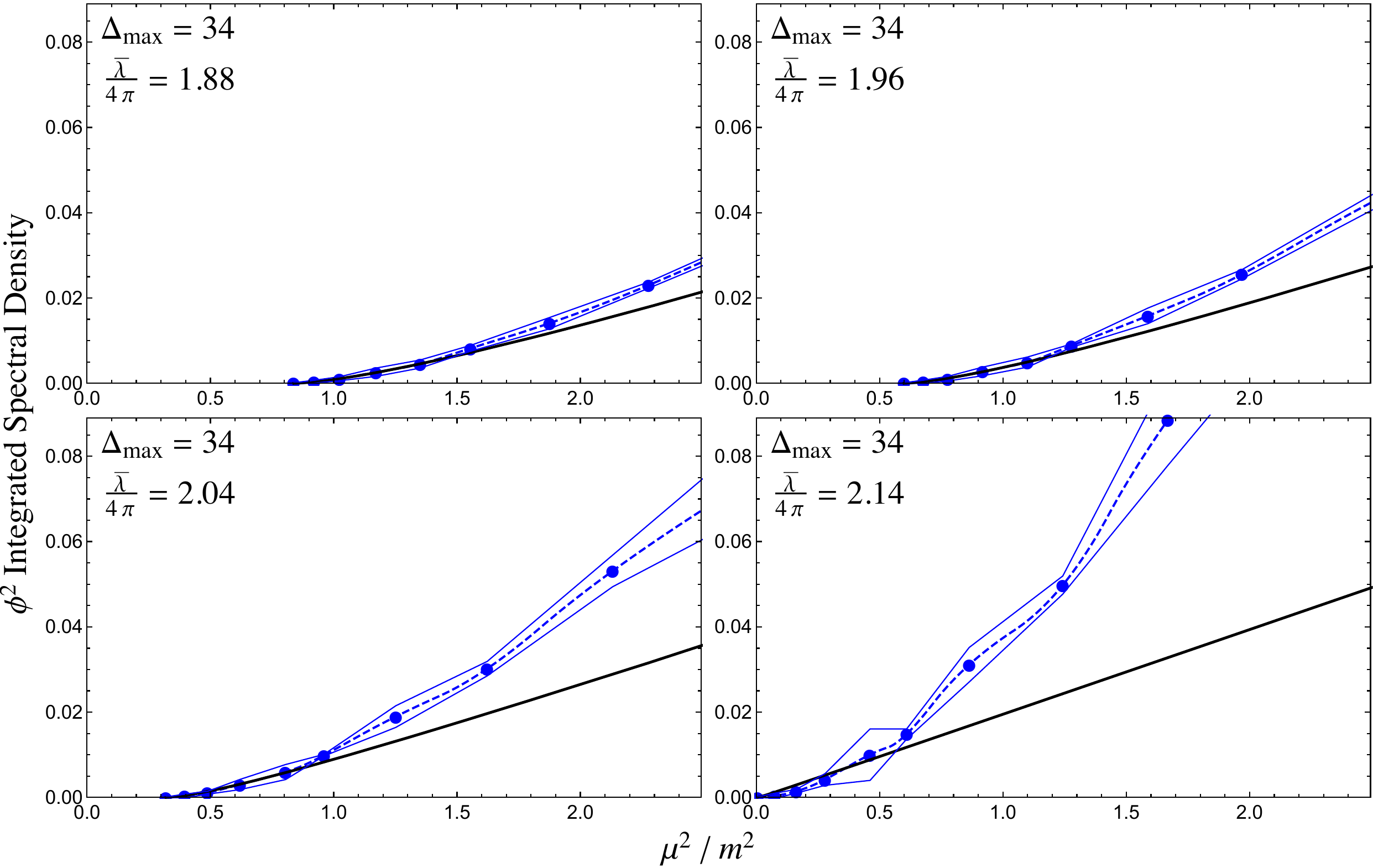}
\caption{Integrated spectral densities for $\phi^2$, for $\Dmax = 34$ and different values of $\lambar$, compared to the Ising model prediction for $\ep$ (black curve). The thin blue lines indicate the magnitude of the difference between these results and those at $\Dmax=30$, providing a rough estimate of the convergence. For reference, the upper right plot corresponds to the same value of the coupling ($\fr{\lambar}{4\pi} = 1.96$) as figures~\ref{fig:PhiNConvergence} and \ref{fig:PhiNZoomOut}.}
\label{fig:EpsilonZoomIn} 
\end{center}
\end{figure}

Figure~\ref{fig:PhiNZoomOut} shows the integrated spectral densities for $\phi^2$, $\phi^4$, and $\phi^6$ (left) and for $\phi$, $\phi^3$, and $\phi^5$ (right). Both plots have $\Dmax=34$ and are at the same couplings as figure~\ref{fig:PhiNConvergence}. Just like for the stess tensor, we have included an envelope surrounding each spectral density whose width indicates the difference between these results and those at $\Dmax=30$. Also, we have again rescaled these results by an overall coefficient, this time such that the very first data points match. In both plots, while the spectral densities are clearly distinct in the UV, they all converge to the same universal behavior in the IR.

This IR universality continues to hold across a range of couplings in the vicinity of $\lambar_*$. As an example, figure~\ref{fig:EpsilonZoomIn} shows the integrated spectral density for $\phi^2$ at different values of $\lambar$, compared with the $\ep$ spectral density. While the results match the theoretical prediction at low energies, the rate of convergence appears to decrease as we push closer to the critical coupling. This is unsurprising, as the resulting spectrum becomes more finely tuned as the mass eigenvalues go to zero, and the truncation results therefore converge more slowly in $\Dmax$, as we saw in section~\ref{sec:Coupling}.

%%%%%%%%%%%%%%%%%%%%%%%%%%%%%%%%%%%%%%%%%%%%%%%%%%%%%%%%%%%%%%%%%%%%%%%%%%%%%

\subsection{$T_{--}$ and the Central Charge}

Finally, we can consider the stress-energy tensor component $T_{--} \equiv (\p_-\phi)^2$. The integrated spectral density for this operator is particularly interesting in 2D, because it corresponds to the spectral representation of the Zamolodchikov $C$-function \cite{Zamolodchikov:1986gt,Cappelli:1990yc,Mussardo:2010mgq},
\be
C(\mu) \equiv \fr{12\pi}{P_-^4} \int_0^{\mu^2} d\mu^{\prime \, 2} \, \rho_{T_{--}}(\mu') = \fr{12\pi}{P_-^4} \sum_{\mu_i \leq \mu} |\<T_{--}(0)|\mu_i\>|^2.
\ee
As is well-known, this function monotonically interpolates between the central charges of the UV and IR fixed points. While we can compute $C(\mu)$ for any coupling $\lambar$, unfortunately near criticality the Ising model prediction is very sensitive to UV corrections, making the comparison with theory more subtle for this particular observable. 

In particular, for RG flows which lead to a non-trivial IR CFT, one could in principle use the spectral density of $T_{--}$ to determine the associated central charge, $c_{\textrm{IR}}$. In practice, if the IR fixed point is fine-tuned, as in $\phi^4$ theory, the resulting truncated spectrum will always have a small but nonzero mass gap. In this case, the $C$-function will flow to the trivial central charge,
\benn
C(\mu) \ra 0 \textrm{ as } \mu \ra 0 \quad (m_\gap \neq 0).
\eenn
If $m_\gap$ is nevertheless sufficiently small compared to the mass scales of the UV theory, the $C$-function will still plateau at $c_{\textrm{IR}}$ before eventually falling to zero as $\mu\ra 0$. Our ability to extract the IR central charge from the $T_{--}$ spectral density is therefore determined by the size of $m_\gap$ relative to the other scales characterizing the RG flow.

\begin{figure}[t!]
\begin{center}
\includegraphics[width=0.7\textwidth]{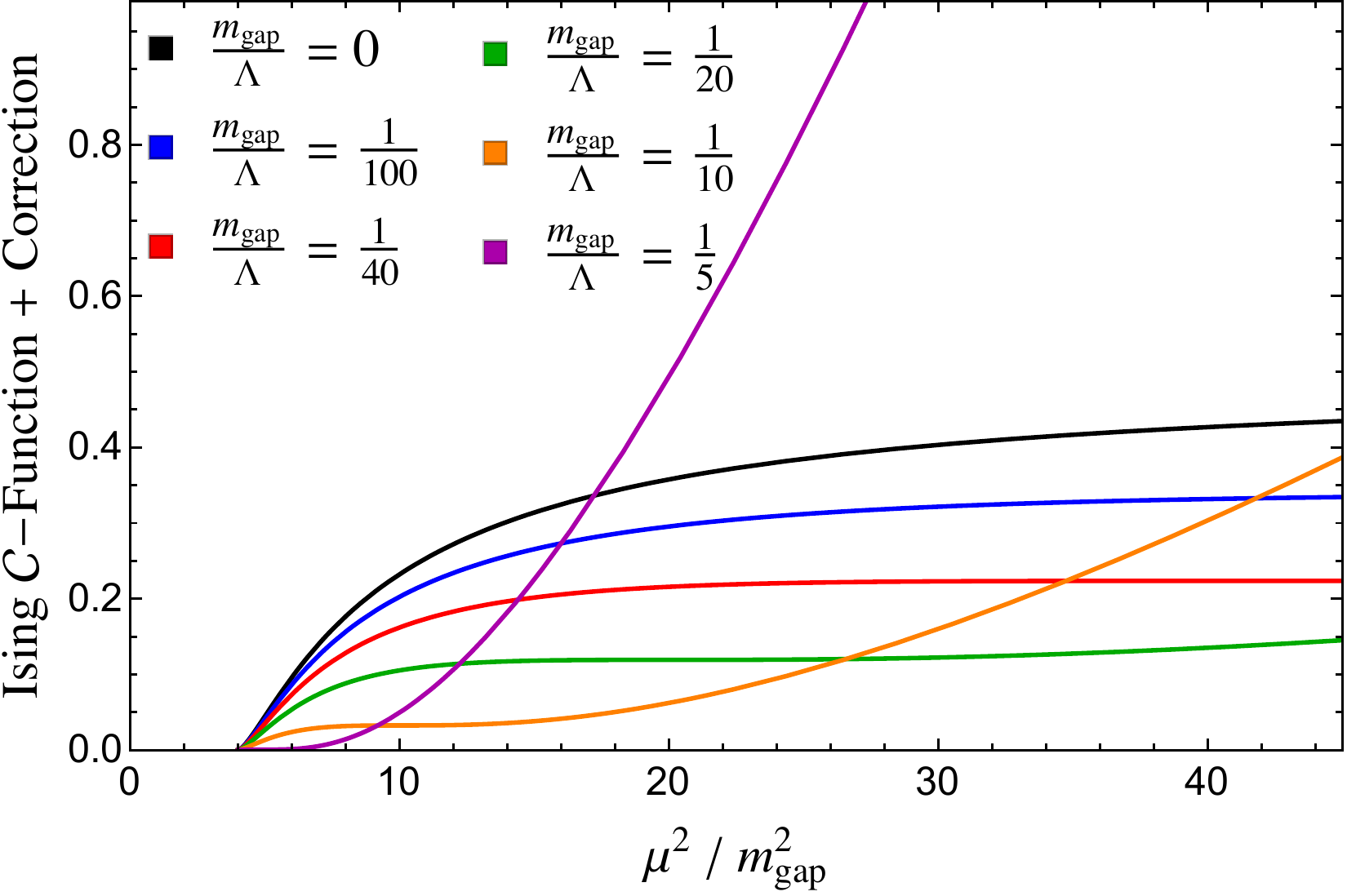}
\caption{Theoretical prediction for the Zamolodchikov $C$-function in the Ising model effective theory, including the correction from the leading irrelevant operator, for different values of $\fr{m_\gap}{\Lambda}$. In the limit $\Lambda \ra \infty$ (black curve), the function levels out and approaches the Ising central charge $c_{\textrm{Ising}} = \half$. For finite values of $\Lambda$, the corrections dramatically alter the function, lowering the plateau and eventually completely eliminating it as $\Lambda \ra m_\gap$.} 
\label{fig:CentralChargeTheory} 
\end{center}
\end{figure}

To be more concrete, the Ising model description of $\phi^4$ theory is merely a low-energy effective theory, with an associated cutoff $\Lambda$ set by the UV parameters $m$ and $\lambda$. The stress-energy tensor, and thus the resulting effective Hamiltonian, receive corrections from higher-dimensional Ising model operators, suppressed by this cutoff,
\be
T_{+-} \approx m_\gap \ep - \fr{\p^2 \ep}{\Lambda} + \cdots
\ee
with the remaining terms suppressed by higher powers of $\Lambda$. Using this effective Ising framework, we can determine the effects of these corrections on spectral densities as a function of the ratio $\fr{m_\gap}{\Lambda}$. For example, if we include the correction due to the leading irrelevant operator $\p^2 \ep$, the prediction for the $T_{--}$ spectral density takes the form
\be
\rho_{T_{--}}(\mu) \approx \fr{P_-^4}{4\pi\mu^4} \sqrt{1-\fr{4m_\gap^2}{\mu^2}} \left(m_\gap^2 - \fr{m_\gap \mu^2}{\Lambda} + \fr{\mu^4}{\Lambda^2}\right).
\label{eq:CFunction}
\ee

Figure~\ref{fig:CentralChargeTheory} shows the resulting Ising model prediction for the $C$-function for different values of $\fr{m_\gap}{\Lambda}$. In the limit $\Lambda \ra \infty$ (black curve), the corrections are negligible and the $C$-function flattens out, allowing us to extract the central charge $c_{\textrm{Ising}} = \half$. However, as we increase $m_\gap$ relative to the cutoff, the corrections rapidly alter the theoretical prediction, such that the plateau is almost completely removed for $\fr{m_\gap}{\Lambda} \gtrsim \fr{1}{10}$.

\begin{figure}[t!]
\begin{center}
\includegraphics[width=\textwidth]{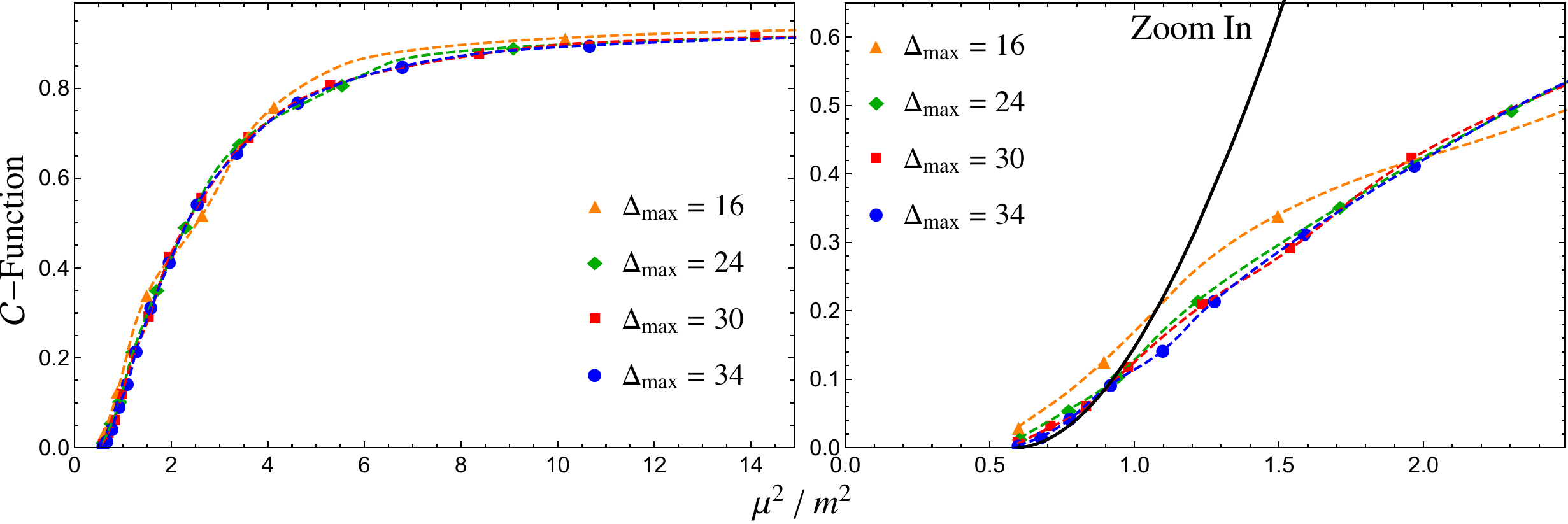}
\caption{Zamolodchikov $C$-function at different values of $\Dmax$. The $\Dmax=34$ results (blue dots) are at $\fr{\lambar}{4\pi} = 1.96$, and the couplings for the remaining results have been chosen such that the mass gap remains fixed. The points are the actual contributions of individual eigenstates to the spectral density, while the dashed lines are interpolations. The right plot is simply a zoomed-in version of the left one, and compares the conformal truncation results to the theoretical IR prediction for the Ising model (black curve), which includes the correction from the leading irrelevant operator (with $\fr{\Lambda}{m} = 1.0$).} 
\label{fig:CentralChargeConvergence} 
\end{center}
\end{figure}

\begin{figure}[t!]
\begin{center}
\includegraphics[width=1\textwidth]{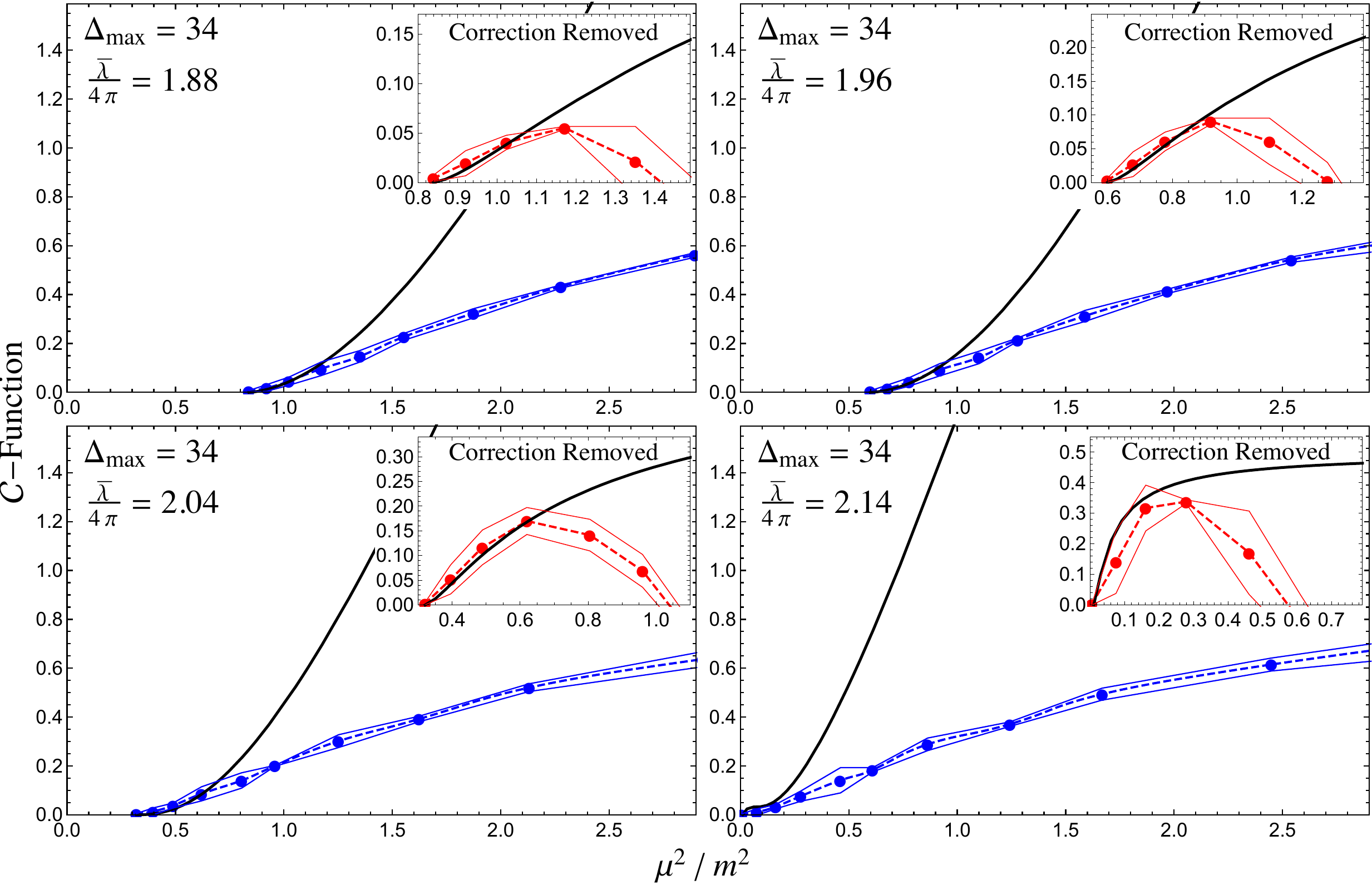}
\caption{Zamolodchikov $C$-function for $\Dmax = 34$ and different values of $\lambar$. The thin lines surrounding the data points indicate the magnitude of the difference between these results and those at $\Dmax=30$, providing a rough estimate of the convergence. Main plots: raw data (blue dots) compared to the Ising model prediction (black curve), which includes the correction from the leading irrelevant operator (with $\fr{\Lambda}{m} = 1.0$). Insets: same data points, but with the expected leading correction removed (red dots), compared with the Ising model prediction (black curve).} 
\label{fig:CentralChargeZoomIn} 
\end{center}
\end{figure}

From this plot, we see that the $C$-function is very sensitive to corrections from UV physics, such that we must set $m_\gap$ far below the cutoff to be able to read off $c_{\textrm{IR}}$ directly. More importantly, though, even the IR behavior of $C(\mu)$ changes dramatically due to UV effects. This suggests that we need to account for these irrelevant operators when comparing our numerical results to the predictions from the Ising model.

We can see this clearly in figure~\ref{fig:CentralChargeConvergence}, which shows our conformal truncation results for the $C$-function at four different values of $\Dmax$. Just like in previous plots, the couplings have been chosen such that the results all have the same mass gap. In the left plot, we see that the results have converged over a wide range of $\mu$, showing the full RG flow from the free scalar central charge $c_{\textrm{UV}} = 1$ at high energies to the trivial value of zero in the IR, with the transition scale roughly corresponding to the coupling $\fr{\lambda}{4\pi}$.

However, there appears to be no plateau in the IR corresponding to $c_{\textrm{Ising}} = \half$, indicating that the effective cutoff $\Lambda$ is not sufficiently large compared to $m_\gap$. We can confirm this by fitting the IR data points with the Ising model prediction, including the correction from the leading irrelevant operator $\p^2 \ep$, as shown in the right plot. The resulting fit yields $\fr{\Lambda}{m} \approx 1.0$, which corresponds to $\fr{m_\gap}{\Lambda} \approx 0.4$.

In order to suppress these corrections and isolate the unperturbed Ising model prediction, we therefore must push the mass gap much lower. However, our truncation to $\Dmax=34$ limits our IR resolution, setting a lower bound on the value of $m_\gap$ we can accurately probe with our numerical results. At this truncation level, we are therefore unable to set $m_\gap$ low enough to ignore these corrections to the $C$-function.

It is important to note that these corrections to the Ising prediction are \emph{not} a result of truncation error. The effective cutoff $\Lambda$ is a physical scale at which the Ising model description of $\phi^4$ theory breaks down, and these corrections are just a consequence of that fact. Truncation effects merely limit the amount of separation we can obtain between $m_\gap$ and $\Lambda$, or equivalently, how close we can get to the critical point.

\begin{figure}[t!]
\begin{center}
\includegraphics[width=0.7\textwidth]{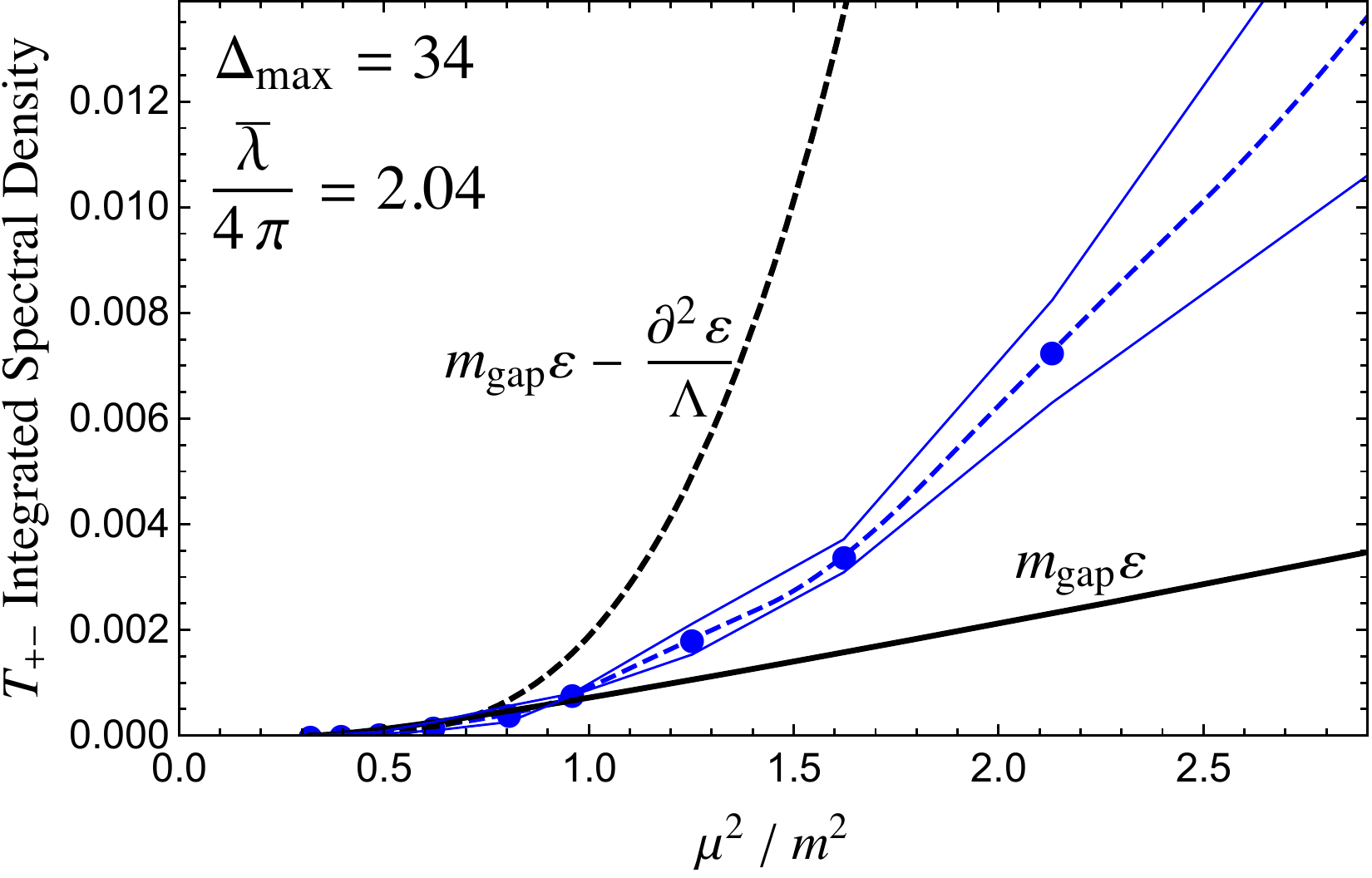}
\caption{Integrated spectral density for $T_{+-}$ at $\Dmax = 34$ and $\fr{\lambar}{4\pi} = 2.04$, compared to the Ising model prediction with (dashed line) and without (solid line) the correction from $\p^2\ep$, with $\fr{\Lambda}{m} = 1.0$. In the IR, the effects from this leading correction are negligible, such that we can safely ignore them. For reference, the numerical results are the same as those in the lower left plot in figure~\ref{fig:TraceZoomIn}.} 
\label{fig:Compare} 
\end{center}
\end{figure}

Figure~\ref{fig:CentralChargeZoomIn} shows the truncation results for the $C$-function for $\Dmax=34$ and multiple values of $\lambar$. In the main plots, we compare these results (blue dots) with the theoretical prediction (Ising + leading correction) with a \emph{fixed} cutoff $\fr{\Lambda}{m} = 1.0$ across the $\lambar$ shown. Even though the mass gap changes significantly as we vary $\lambar$, the IR data points continue to be well-described by eq.~\eqref{eq:CFunction}. The insets in these plots confirm this agreement, showing the truncation results for $C(\mu)$ with the expected corrections from $\p^2 \ep$ removed (red dots). In the IR, these modified results now match the original Ising model predictions (i.e.~without any corrections from irrelevant operators), again indicating that our truncation results are correctly reproducing the effects due to the cutoff $\Lambda$.

One obvious question is whether the corrections due to irrelevant operators also have a significant effect on the integrated spectral densities for $T_{+-}$ and $\phi^n$. After all, in the previous subsections we completely ignored these effects when comparing our truncation results with theoretical predictions. Fortunately, unlike for $T_{--}$, the corrections to those spectral densities are negligible in the IR. As an example, figure~\ref{fig:Compare} shows the theoretical prediction for the $T_{+-}$ integrated spectral density, both with and without the leading correction from $\p^2\ep$, compared with the conformal truncation results at $\Dmax = 34$ and $\fr{\lambar}{4\pi} = 2.04$. In the IR, the two theoretical predictions agree, indicating that we can safely ignore the corrections from higher-dimensional operators when comparing with our numerical results.

%%%%%%%%%%%%%%%%%%%%%%%%%%%%%%%%%%%%%%%%%%%%%%%%%%%%%%%%%%%%%%%%%%%%%%%%%%%%%
%%%%%%%%%%%%%%%%%%%%%%%%%%%%%%%%%%%%%%%%%%%%%%%%%%%%%%%%%%%%%%%%%%%%%%%%%%%%%

\section{Discussion}
\label{sec:Discussion}

Conformal truncation, which we introduced in~\cite{Truncation}, is a new method for performing non-perturbative computations in strongly-coupled QFTs. Unlike other numerical methods, it is formulated in Lorentzian signature and infinite volume and consequently can be used to compute real-time, continuum correlation functions. In this work, we have used conformal truncation to specifically calculate \KL spectral densities, which are equivalent to two-point functions. To the best of our knowledge, the results presented here constitute the first computation of non-perturbative spectral densities in 2D $\phi^4$ theory.

Our main goal has been to check these spectral densities against known analytic results in the IR limit, as a test of our conformal truncation method. As such, we have focused on values of the coupling, $\lambar$, near the critical point, where we know the IR theory is described by the 2D Ising model. In section~\ref{sec:Ising}, we demonstrated that in this regime the spectral densities for several different operators match known Ising spectral densities at low energies, providing a fully non-perturbative check of conformal truncation. 

It is worth emphasizing two things. First, our truncation results extend well beyond the deep IR regime described by the Ising model. As we have shown, the spectral densities converge rapidly in $\Dmax$ over a wide range of mass scales, $\mu^2$, providing the full RG flow of the corresponding operators. These are novel predictions for non-perturbative two-point functions in $\phi^4$ theory. Second, as we vary $\Dmax$, the resulting spectral densities are built from the IR up. That is, the convergence of the spectral densities starts at low mass scales and then extends to larger $\mu^2$ with increasing $\Dmax$. This is evident, for example, in the convergence plots in figure~\ref{fig:TraceConvergence}, where even $\Dmax=16$ correctly reproduces the IR. This capacity to preferentially access IR physics is a useful feature of conformal truncation. 

Our analysis has taught us some general lessons about conformal truncation. One clear lesson is that conformal truncation becomes less efficient as we increase the separation between the bare parameters in the UV Hamiltonian and emergent IR scales like $m_{\gap}$. From a computational perspective, this is simply because a small value for $m_{\gap}$ is the result of fine-tuned cancellations between UV basis states. As $m_\gap$ decreases, the IR results thus become increasingly sensitive to small corrections from operators with large conformal Casimir. This is most pronounced at a critical point, where $m_{\gap}$ vanishes, and explains why in figure~\ref{fig:LowestEvalsExtrapolate} the error bars increase as we approach criticality. We can also see this behavior in the various spectral density results, where the convergence slows as we tune $m_\gap \ra 0$. This inability to fully reach criticality at finite $\Dmax$ thus makes it difficult to extract observables like critical exponents and central charges using conformal truncation.

Another important lesson can be drawn by comparing the convergence of our results in sections~\ref{sec:Coupling} and~\ref{sec:Ising}. In section~\ref{sec:Ising}, we found that the spectral densities converged quite rapidly if we held the IR observable $m_\gap$ fixed. We can contrast this with the mass spectrum results in section~\ref{sec:Coupling}, where we instead held the UV parameter $\lambar$ fixed. Even visually, it is clear that the latter results converge much more slowly than the former ones. This is perhaps unsurprising, as mapping precisely between the mass gap and $\lambar$ requires reconstructing the \emph{entire} RG flow. Because conformal truncation constructs observables from the IR up, it is thus much more efficient to study low-energy physics directly in terms of IR parameters, rather than UV ones.

Perhaps one way to summarize these observations is that conformal truncation appears to be truly complementary to existing numerical methods. While most methods excel at computing critical observables, conformal truncation is better at studying full RG flows. Conformal truncation can thus deliver something new: real-time, infinite-volume correlation functions computable efficiently in $\Dmax$.  

Finally, it is worth commenting that in the space of CFTs, there is a precise sense in which the 2D free scalar CFT is actually the \emph{least} efficient setting for conformal truncation. In 2D free field theory, a primary operator $\Ocal$ only corresponds to a single state in the Hilbert space, as we discussed in section~\ref{sec:Model}. For more general theories, each operator $\Ocal$ gives rise to a continuum of states, parameterized by the invariant mass $\mu$. Equivalently, we can think of these additional states as being created by the descendant operators $P^{2n} \Ocal$. Computationally, constructing primary operators is expensive, while including additional descendants is quite cheap. The 2D free scalar CFT thus has the least return in terms of the number of basis states obtained with a given computational power. As a concrete point of comparison, in~\cite{Truncation} we considered the 3D free scalar CFT as a starting point for studying the $O(N)$ model in the limit $N\rightarrow\infty$. Using conformal truncation, we were able to reproduce the IR spectral density of the singlet operator $\phivec$, in roughly equivalent detail to the results presented here, using just $\Dmax\sim 5$. The key difference in that work is that we were able to increase the size of our truncated basis with descendants. For this reason, we are optimistic about the capabilities of conformal truncation moving forward to other theories. 

Looking ahead, there are several exciting applications of conformal truncation to pursue:
\begin{enumerate}
\item[{\bf 1)}] {\bf 2D $\phi^4$ theory -- continued} 

In this work, we have only studied the symmetry-preserving phase of $\phi^4$ theory, focusing particularly on couplings below $\lambar_*$. However, as mentioned above, conformal truncation yields results for any $\lambar$, so a natural next step is to proceed to the symmetry-broken phase. There is reason to believe that, despite the triviality of the vacuum, spontaneous symmetry-breaking is detectable even in lightcone quantization~\cite{Rozowsky:2000gy}, and some initial work has been done in~\cite{Chakrabarti:2003tc,Chakrabarti:2003ha,Chakrabarti:2005zy}. It would thus be illuminating to study the behavior of spectral densities in the symmetry-broken phase.

On a different note, it would also be interesting to further study the map between lightcone and equal-time quantization. In particular, it would be instructive to use the prescription presented in~\cite{Burkardt:2016ffk} to see if one can explicitly map our results to those done in equal-time. This would allow us to compare the value of the critical coupling across the two quantization schemes.
 
\item[{\bf 2)}] {\bf 2D Ising model} 

In this work, we merely used the 2D Ising model to check our method, making use of the fact that an $\ep$ (temperature) deformation is integrable and can be treated analytically. However, it would be fascinating to use conformal truncation to systematically study the more general case of deforming the 2D Ising model by both $\ep$ and $\sigma$, which corresponds to the Ising model at $T\neq T_c$ in a magnetic field. Correlation functions in this theory are not known, and conformal truncation could potentially be used to make novel predictions. There are two strategies for doing this. 

The first strategy is to again consider $\phi^4$ theory, but now with an additional $\mathbb{Z}_2$-odd $\phi^3$ deformation. Our results here have confirmed that $\phi^3$ flows to $\sigma$ near criticality, so adding this interaction is equivalent in the IR to deforming the Ising model by a magnetic field. The advantage of this approach is that the UV CFT is still free scalar field theory, so the basis of primary operators is the same one we used in this work. The disadvantage is that flowing all the way from free field theory is an inefficient use of computational power, using thousands of UV operators to reproduce only a handful of Ising model states. 

The second strategy, which we suspect is much more efficient, is to apply conformal truncation directly to the 2D Ising CFT. Indeed, conformal truncation can be initiated from any UV CFT where operator scaling dimensions and OPE coefficients are known up to a desired truncation level. Since the 2D Ising CFT is a minimal model where all of this data is known, it seems more sensible to start directly from this CFT and use conformal truncation to construct and diagonalize the Hamiltonian created by the $\ep$ and $\s$ deformations.

\item[{\bf 3)}] {\bf 3D Ising model} 

Another important feature of conformal truncation is that it can be applied in any number of spacetime dimensions. Thus, a natural goal is to use this method to study the 3D Ising model, about which much less is known than its 2D counterpart. As in 2D, there are two approaches to studying deformations of the 3D Ising model: starting in scalar field theory and flowing to the vicinity of the Ising critical point, which we plan to consider in future work,\footnote{See~\cite{Hogervorst:2014rta} for an alternative proposal for applying Hamiltonian truncation to $\phi^4$ theory in higher dimensions.} or starting directly from the Ising CFT and deforming it. 

The advantage of starting from free field theory is always that we know operator dimensions and OPE coefficients, which are the necessary ingredients for conformal truncation. By comparison, this data is difficult to obtain in the 3D Ising CFT. A direct application of conformal truncation to the 3D Ising model would require us to know the operator content and OPE coefficients up to a desired $\Dmax$. Over the past several years, there has been remarkable progress in pinning down 3D Ising data using the conformal bootstrap and related techniques~\cite{El-Showk:2014dwa,Gliozzi:2014jsa,Kos:2016ysd,Simmons-Duffin:2016wlq}. It may turn out that these techniques can provide the CFT data needed to subsequently initiate truncation studies directly around the 3D Ising model. More generally, conformal truncation applications provide an immediate incentive for trying to compute scaling dimensions and OPE coefficients in known CFTs. 

\end{enumerate}

%%%%%%%%%%%%%%%%%%%%%%%%%%%%%%%%%%%%%%%%%%%%%%%%%%%%%%%%%%%%%%%%%%%%%%%%%%%%%
%%%%%%%%%%%%%%%%%%%%%%%%%%%%%%%%%%%%%%%%%%%%%%%%%%%%%%%%%%%%%%%%%%%%%%%%%%%%%

\section*{Acknowledgments}    

It is a pleasure to thank Rich Brower, Chris Brust, Liam Fitzpatrick, Matthijs Hogervorst, Guoan Hu, Jared Kaplan, Markus Luty, Gustavo Marques Tavares, Juli\'{a}n Mu\~{n}oz, Slava Rychkov, Balt van Rees, and Lorenzo Vitale for valuable discussions. We are especially grateful to Liam and Lorenzo for helpful comments on the draft. This work was supported in part by DOE grant DE-SC0015845. NA was supported by a Johns Hopkins University Catalyst Award and Sloan Foundation fellowship BR2014-031. VG holds a postdoctoral fellowship from the Natural Science and Engineering Research Council of Canada. ZK acknowledges support from the DARPA, YFA Grant D15AP00108. We would also like to thank the Galileo Galilei Institute for Theoretical Physics for hospitality while this work was completed.

%%%%%%%%%%%%%%%%%%%%%%%%%%%%%%%%%%%%%%%%%%%%%%%%%%%%%%%%%%%%%%%%%%%%%%%%%%%%%
%%%%%%%%%%%%%%%%%%%%%%%%%%%%%%%%%%%%%%%%%%%%%%%%%%%%%%%%%%%%%%%%%%%%%%%%%%%%%
%%%%%%%%%%%%%%%%%%%%%%%%%%%%%%%%%%%%%%%%%%%%%%%%%%%%%%%%%%%%%%%%%%%%%%%%%%%%%

\appendix

\section{Basis of Casimir Eigenstates}
\label{app:Basis}

Our basis consists of total momentum eigenstates built from local operators in the UV CFT,\footnote{Note that we have suppressed the indices on the coordinates and momentum, with the understanding that all indices are ``$-$''.}
\be
|\Ccal,P\> \equiv \int dx \, e^{-iPx} \Ocal(x)|0\>,
\label{eq:BasisDef}
\ee
with the normalization convention
\be
\<\Ccal,P|\Ccal',P'\> = 2P (2\pi) \de(P-P') \, \de_{\Ocal\Ocal'}.
\ee
As our CFT is free scalar field theory, the operators can be written in terms of derivatives acting on the scalar field $\phi$,
\be
\Ocal(x) = \sum_{\kvec} C^\Ocal_{\kvec} \, \p^{k_1}\phi(x) \cdots \p^{k_n}\phi(x) \equiv \sum_{\kvec} C^\Ocal_{\kvec} \, \p^{\kvec}\phi(x),
\ee
where we have introduced the useful shorthand
\be
\p^{\kvec} \phi \equiv \p^{k_1} \phi \cdots \p^{k_n} \phi.
\ee

We specifically need to find the linear combinations that correspond to \emph{primary operators}, which are annihilated by the special conformal generator $K_\mu$ and create eigenstates of the conformal quadratic Casimir $\Ccal$,
\be
\comm{K_\mu}{\Ocal(0)} = 0, \qquad \comm{\Ccal}{\Ocal(0)} = \big( \De(\De-2) + \ell^2 \big) \Ocal(0).
\label{eq:PrimaryDef}
\ee
There are two ways to obtain the set of primary operators. The first, more direct method is to simply construct linear combinations which satisfy eq.~\eqref{eq:PrimaryDef} by brute force. The conformal Casimir and special conformal generator can be written as operators acting on the space of ``monomials'' $\p^{\kvec} \phi$, such that constructing primary operators is equivalent to simply organizing the null space of $K_\mu$ into eigenstates of $\Ccal$.

The second method, which we use in this work, is to first construct a basis of primary operators built from distinguishable particles, then symmetrize with respect to particle number. In other words, we first find operators of the form
\benn
\Ocal(x) = \sum_{\sigvec} C^\Ocal_{\sigvec} \, \p^{\sigma_1}\phi_1(x) \cdots \p^{\sigma_n}\phi_n(x),
\eenn
with $n$ distinct fields $\phi_i$. We can then remove the labels on $\phi_i$ to obtain primary operators built from a single scalar field. The advantage of this approach is that the restriction to primary operators and organization into Casimir eigenstates is much simpler for states with distinguishable particles.

The Casimir eigenstates created by these operators can be expressed in terms of $n$-particle Fock space states. Each operator $\Ocal(x)$ maps to a corresponding ``wavefunction'' $F_\Ocal(p)$, defined as the overlap
\be
F_\Ocal(p_1,\ldots,p_n) \equiv \<p_1,\ldots,p_n|\Ocal(0)\>,
\ee
allowing us to rewrite the corresponding basis states as
\be
|\Ccal,P\> = \fr{1}{n!} \int \fr{dp_1 \cdots dp_n}{(2\pi)^n 2p_1 \cdots 2p_n} (2\pi) \de\Big(P - \sum_i p_i\Big) F_\Ocal(p) |p_1,\ldots,p_n\>.
\label{eq:FockSpaceStates}
\ee
The advantage of working with momentum space wavefunctions is that this representation \emph{automatically} restricts our basis to primary operators. This simplification occurs because descendants are created by acting with overall derivatives on primary operators, which in terms of Fock space states simply corresponds to multiplying the wavefunction by a constant,
\benn
\p^k \Ocal(x) \ra (p_1 + \cdots + p_n)^k F_\Ocal(p) = P^k F_\Ocal(p).
\eenn

Now that we have restricted our basis to primary operators, we can use the methods of \cite{Truncation} to solve for the complete set of eigenfunctions of the conformal quadratic Casimir,
\be
\Ccal = -D^2 - \half (P_\mu K^\mu + K_\mu P^\mu) + \half L_{\mu\nu} L^{\mu\nu},
\ee
which can be written as the momentum space differential operator,
\be
\Ccal = -2 \sum_{i<j} p_i p_j \left( \fr{\p}{\p p_i} - \fr{\p}{\p p_j} \right)^2.
\ee
The resulting Casimir eigenfunctions are multivariate Jacobi polynomials, parameterized by the set of indices $\Lvec \equiv (\ell_1,\ldots,\ell_{n-1})$,
\be
\boxed{F_{\Lvec}(p) = p_1 \cdots p_n \prod_{i=1}^{n-1} |p|_{i+1}^{\ell_i} \, P^{(2|\ell|_{i-1}+2i-1,1)}_{\ell_i} \left( \fr{p_{i+1}-|p|_i}{|p|_{i+1}} \right),}
\label{eq:MultiJacDef}
\ee
where we have ignored the overall normalization coefficient and defined
\be
|p|_i \equiv \sum_{j=1}^i p_j.
\ee

These Casimir eigenfunctions can be converted back into local operators simply by making the identification
\benn
p_i^{k_i} \ra \p^{k_i} \phi_i.
\eenn
We can see this more concretely by expanding the wavefunctions into sums of monomials, then using the monomial coefficients to construct the corresponding operator,
\be
F_{\Lvec}(p) = \sum_{\sigvec} C^{\Lvec}_{\sigvec} \, p_1^{\sigma_1} \cdots p_n^{\sigma_n} \ra \Ocal_{\Lvec}(x) = \sum_{\sigvec} C^{\Lvec}_{\sigvec} \, \p^{\sigma_1}\phi_1(x) \cdots \p^{\sigma_n}\phi_n(x).
\ee
Finally, we can remove the indices on the individual scalar fields to obtain the resulting primary operator
\be
\Ocal_{\Lvec}(x) = \sum_{\sigvec} C^{\Lvec}_{\sigvec} \, \p^{\sigma_1}\phi(x) \cdots \p^{\sigma_n}\phi(x) = \sum_{\kvec} \Bigg( \sum_{\sigvec \in \textrm{perm}(\kvec)} C^{\Lvec}_{\sigvec} \Bigg) \p^{\kvec}\phi(x).
\ee

As a simple example, let's consider the two-particle Casimir eigenfunction with $\ell=2$,
\benn
F_2(p_1,p_2) = p_1 p_2 \, (p_1 + p_2)^{2} \, P^{(1,1)}_{2} \left( \fr{p_2-p_1}{p_1+p_2} \right) = 3 p_1^3 p_2 + 3p_1 p_2^3 - 9 p_1^2 p_2^2.
\eenn
This polynomial can be used to construct an operator built from two distinct fields,
\benn
F_2(p) \ra \Ocal_2 = 3 \p^3\phi_1 \p \phi_2 + 3 \p\phi_1 \p^3\phi_2 - 9 \p^2 \phi_1 \p^2 \phi_2.
\eenn
We can then replace $\phi_{1,2} \ra \phi$ and collect together similar terms to obtain the final operator
\be
\Ocal_2 = 6 \p^3\phi \p\phi - 9 (\p^2\phi)^2.
\ee

We thus have a straightforward procedure for constructing the basis of Casimir eigenstates. Starting with the polynomials in eq.~\eqref{eq:MultiJacDef}, we can convert each wavefunction into a corresponding primary operator built from $n$ distinct fields. We can then obtain operators built from a single scalar field by simply replacing $\phi_i \ra \phi$.

An alternative approach would be to first symmetrize the momentum space wavefunctions with respect to particle number, then convert the resulting symmetric polynomials into operators built from a single scalar field. However, this symmetrization procedure is much simpler when implemented at the level of operators. Our approach therefore capitalizes on the relative advantages of both representations of the basis. Working in momentum space trivializes the restriction to primary operators, while converting back to operators in position space trivializes the process of symmetrization.

Because of this need to symmetrize, the set of eigenfunctions in~\eqref{eq:MultiJacDef} is overcomplete, which means that multiple polynomials will map to the same final operator (or to linearly dependent combinations of operators). In practice, we therefore only need to use a subset of the Casimir eigenfunctions to span the space of primary operators, using Gram-Schmidt to find the orthogonal linear combinations. A more detailed discussion of this process, as well as its generalization to higher dimensions, will be presented in future work \cite{FutureUs}.

In \cite{Burkardt:2016ffk} (based on initial work in \cite{Chabysheva:2013oka,Chabysheva:2014rra}), Burkardt et al.~considered a basis of Fock space states weighted by symmetric polynomials in momentum space. They then truncated this basis by setting a separate maximum degree for the polynomials in each $n$-particle sector. The resulting basis states are linear combinations of the Casimir eigenstates we use in this work, such that their truncation scheme is equivalent to setting a different value of $\Dmax$ for each particle number in our basis. One can see this explicitly by either computing the wavefunctions $F_\Ocal(p)$ of our final basis of Casimir eigenstates or converting the symmetric polynomials used in~\cite{Burkardt:2016ffk} into local operators built from $\phi$. In practice, we find that working in terms of operators, rather than polynomials, greatly simplifies the construction and orthogonalization of the basis.

%%%%%%%%%%%%%%%%%%%%%%%%%%%%%%%%%%%%%%%%%%%%%%%%%%%%%%%%%%%%%%%%%%%%%%%%%%%%%
%%%%%%%%%%%%%%%%%%%%%%%%%%%%%%%%%%%%%%%%%%%%%%%%%%%%%%%%%%%%%%%%%%%%%%%%%%%%%

\section{Matrix Elements and Operator Overlaps}
\label{app:Matrix}

In this appendix, we use our basis of Casimir eigenstates to compute matrix elements for the invariant mass operator $M^2$. While we are technically only interested in the matrix elements associated with primary operators, in practice it is simpler to first evaluate the expressions for individual ``monomials,''
\be
|\p^{\kvec}\phi,P\> \equiv \int dx \, e^{-iPx} \p^{\kvec}\phi(x)|0\>,
\label{eq:monstate}
\ee
which can then be combined to form matrix elements for the primary operators
\be
|\Ccal,P\> = \sum_{\kvec} C^\Ocal_{\kvec} |\p^{\kvec}\phi,P\>.
\ee
These monomial matrix elements take the general form
\be
\<\p^{\kvec}\phi,P| M^2 |\p^{\kvec'}\phi,P'\> = 2P (2\pi) \de(P-P') \, \Mcal_{\kvec\kvec'}.
\ee
For the rest of this discussion, we will focus only on the dynamical piece $\Mcal_{\kvec\kvec'}$, suppressing the momentum-conserving kinematic factor. Note that, because our states are lightcone momentum eigenstates, the matrix elements can be further simplified to
\be
\Mcal_{\kvec\kvec'} \equiv \<\p^{\kvec}\phi| M^2 |\p^{\kvec'}\phi\> = 2P\<\p^{\kvec}\phi|P_+|\p^{\kvec'}\phi\>.
\ee
Constructing $\Mcal_{\kvec\kvec'}$ is thus equivalent to calculating the matrix elements for the lightcone Hamiltonian $P_+$.

In this work, we specifically consider the scalar field theory arising from the Lagrangian
\be
\Lcal = \half \p_\mu \phi \p^\mu \phi - \half m^2 \phi^2 - \fr{1}{4!} \lambda \phi^4,
\ee
with the corresponding lightcone Hamiltonian
\be
P_+ = \int dx \left( \half m^2 \phi^2 + \fr{1}{4!} \lambda \phi^4 \right).
\ee
Note that the Hamiltonian does not receive any contributions from the kinetic term. This is due to the fact that our basis states are only built from the right-moving operator $\p\phi$, such that every state in the original CFT has invariant mass $\mu^2 = 0$.

The resulting Hamiltonian matrix elements are simply Fourier transforms of CFT three-point functions involving $\phi^2$ and $\phi^4$. It will therefore be useful to evaluate the general integral\footnote{For simplicity, from now on we will suppress any overall factors of $i$, as these cancel in the final matrix elements.}
\be
\begin{split}
&\int dx \, dy \, dz \, e^{i(Px - P'z)} \fr{1}{(x-y)^a(y-z)^b(x-z)^c} \\
&\qquad \qquad = \fr{2\pi^2 P^{a+b+c-3} \G(a+b-1)}{\G(a)\G(b)\G(a+b+c-1)} \cdot 2P (2\pi)\de(P-P').
\end{split}
\label{eq:3ptIntegral}
\ee

%%%%%%%%%%%%%%%%%%%%%%%%%%%%%%%%%%%%%%%%%%%%%%%%%%%%%%%%%%%%%%%%%%%%%%%%%%%%%

\subsection{Mass Term}

Let us first consider the mass term, which in lightcone quantization preserves particle number. We therefore only need to compute the $n \ra n$ matrix element
\be
\<\p^{\kvec}\phi,P| M^2 |\p^{\kvec'}\phi,P'\> = \fr{m^2}{2} \int dx \, dy \, dz \, e^{i(Px - P'z)} \<\p^{\kvec}\phi(x) \phi^2(y) \p^{\kvec'}\phi(z)\>.
\ee
The three-point function in the integrand can be written as a sum of Wick contractions,
\be
\<\p^{\kvec}\phi(x) \phi^2(y) \p^{\kvec'}\phi(z)\> = \sum_{\substack{k_i \in \kvec \\ k'_j \in \kvec'}} \<\p^{k_i} \phi(x) \phi^2(y) \p^{k'_j} \phi(z)\> \<\p^{\kvec/k_i}\phi(x) \, \p^{\kvec'/k'_j} \phi(z)\>,
\ee
where $\kvec/k_i$ simply indicates the vector obtained by removing $k_i$ from $\kvec$.

Each term in this sum cleanly factorizes into a product of interacting and spectating correlation functions. The piece involving the spectating particles can also be computed from Wick contractions
\be
\<\p^{\kvec/k_i}\phi(x) \p^{\kvec'/k'_j}\phi(z)\> = \fr{A_{\kvec/k_i,\kvec'/k'_j}}{(4\pi)^{n-1} (x-z)^{\De+\De'-k_i-k'_j}},
\ee
where we have defined the Wick contraction coefficient
\be
A_{\kvec\kvec'} \equiv \sum_{\textrm{pairs}} \prod_{i,j} \G(k_i+k'_j).
\ee
The remaining interacting piece can be easily calculated to obtain
\be
\<\p^{k_i}\phi(x) \phi^2(y) \p^{k'_j}\phi(z)\> = 2 \cdot \fr{\G(k_i)\G(k'_j)}{(4\pi)^2 (x-y)^{k_i} (y-z)^{k'_j}}.
\ee

We can combine these three-point functions with the general integral in eq.~\eqref{eq:3ptIntegral} to obtain the final matrix elements
\be
\boxed{\Mcal^{(m)}_{\kvec\kvec'} = m^2 N_{\kvec\kvec'} \sum_{\substack{k_i \in \kvec \\ k'_j \in \kvec'}} \G(k_i+k'_j-1) A_{\kvec/k_i,\kvec'/k'_j},}
\ee
where we have simplified the expression by introducing the overall coefficient
\be
N_{\kvec\kvec'} \equiv \fr{P^{\De+\De'-2}}{4^n \pi^{n-1} \G(\De+\De'-1)}.
\ee

%%%%%%%%%%%%%%%%%%%%%%%%%%%%%%%%%%%%%%%%%%%%%%%%%%%%%%%%%%%%%%%%%%%%%%%%%%%%%

\subsection{Interaction Terms}

We now turn to the contribution from the quartic interaction, which has two distinct types of matrix elements. The first preserves particle number, and the associated three-point function is similar to that of the mass term, though now there are two particles from each state participating in the interaction,
\be
\<\p^{\kvec}\phi(x) \phi^4(y) \p^{\kvec'}\phi(z)\> = \sum_{\substack{k_{i,j} \in \kvec \\ k'_{r,s} \in \kvec'}} \<\p^{k_{i,j}} \phi (x) \phi^4(y) \p^{k'_{r,s}} \phi(z)\> \<\p^{\kvec/k_{i,j}}\phi(x) \, \p^{\kvec'/k'_{r,s}} \phi(z)\>.
\ee
We therefore just need to compute the correlation function
\be
\<\p^{k_{i,j}} \phi (x) \phi^4(y) \p^{k'_{r,s}} \phi(z)\> = 4! \cdot \fr{ \G(k_i)\G(k_j)\G(k'_r)\G(k'_s)}{(4\pi)^4 (x-y)^{k_i+k_j} (y-z)^{k'_r+k'_s}}.
\ee
Using the same approach as the mass term, we then obtain the $n\ra n$ matrix elements
\be
\boxed{\Mcal^{(n\ra n)}_{\kvec\kvec'} = \fr{g}{4\pi} N_{\kvec\kvec'} \sum_{\substack{k_{i,j} \in \kvec \\ k'_{r,s} \in \kvec'}} \fr{\G(k_i)\G(k_j)\G(k'_r)\G(k'_s)\G(k_i+k_j+k'_r+k'_s-1)}{\G(k_i+k_j)\G(k'_r+k'_s)} A_{\kvec/k_{i,j},\kvec'/k'_{r,s}}.}
\ee

The second type of matrix element changes particle number by two, so we also need to consider the correlation function
\be
\<\p^{k_i} \phi (x) \phi^4(y) \p^{k'_{r,s,t}} \phi(z)\> = 4! \cdot \fr{\G(k_i)\G(k'_r)\G(k'_s)\G(k'_t)}{(4\pi)^4 (x-y)^{k_i} (y-z)^{k'_r+k'_s+k'_t}}.
\ee
We then obtain the resulting $n\ra n+2$ matrix elements
\be
\boxed{\Mcal^{(n\ra n+2)}_{\kvec\kvec'} = \fr{g}{4\pi} N_{\kvec\kvec'} \sum_{\substack{k_i \in \kvec \\ k'_{r,s,t} \in \kvec'}} \fr{\G(k'_r)\G(k'_s)\G(k'_t)\G(k_i+k'_r+k'_s+k'_t-1)}{\G(k'_r+k'_s+k'_t)} A_{\kvec/k_i,\kvec'/k'_{r,s,t}}.}
\ee

%%%%%%%%%%%%%%%%%%%%%%%%%%%%%%%%%%%%%%%%%%%%%%%%%%%%%%%%%%%%%%%%%%%%%%%%%%%%%

\subsection{Overlap of $\phi^n$ with Basis States}

Using the matrix elements from this appendix, we can construct and diagonalize the truncated matrix $M^2$. The resulting approximate mass eigenstates can then be used to compute the integrated spectral density for any local operator $\Ocal(x)$, defined in eq.~(\ref{eq:I}). The approximate eigenstates $|\mu_i\>$ are expressed in the UV basis of conformal Casimir eigenstates, so to obtain the integrated spectral density, we need to first compute the overlap of $\Ocal(x)$ with the original basis states. Much like with the matrix elements, in practice it is simpler to evaluate the overlap with the monomial states $|\p^{\kvec}\phi,P\>$, then arrange them into states created by primary operators,
\be
\<\Ocal(0)|\Ccal',P\> = \sum_{\kvec} C^{\Ocal'}_{\kvec} \, \<\Ocal(0)|\p^{\kvec}\phi,P\>.
\ee

In this work, we are specifically interested in the spectral densities associated with the scalar operators $\phi^n$. The corresponding overlap is just the Fourier transform
\be
\<\phi^n(0)|\p^{\kvec}\phi,P\> = \int dx \, e^{iPx} \<\p^{\kvec}\phi(x) \phi^n(0)\>.
\ee
We therefore need to compute the two-point function,
\be
\<\p^{\kvec}\phi(x) \phi^n(0)\> = \fr{n!\G(k_1)\cdots\G(k_n)}{(4\pi)^n x^\De},
\ee
which we can use to obtain the final overlap
\be
\boxed{
\<\phi^n(0)|\p^{\kvec}\phi,P\> = \fr{n! P^{\De-1} \G(k_1) \cdots \G(k_n)}{2^{2n-1}\pi^{n-1}\G(\De)}.
}
\ee

%%%%%%%%%%%%%%%%%%%%%%%%%%%%%%%%%%%%%%%%%%%%%%%%%%%%%%%%%%%%%%%%%%%%%%%%%%%%%
%%%%%%%%%%%%%%%%%%%%%%%%%%%%%%%%%%%%%%%%%%%%%%%%%%%%%%%%%%%%%%%%%%%%%%%%%%%%%

\section{Decoupling of Higher-Dimensional Operators}
\label{app:Decoupling}

In this appendix, we use the asymptotic behavior of the $M^2$ matrix elements to study the convergence of our conformal truncation method. In particular, we would like to understand how both the IR cutoff and corrections to low-energy observables behave as $\Dmax \ra \infty$. Our analysis here is largely based on~\cite{Rychkov:2014eea,Rychkov:2015vap,Hogervorst:2014rta}.

In conformal truncation (or any truncation prescription), we divide the Hilbert space of a given QFT into two sectors,
\be
\Hcal = \Hcal_L \oplus \Hcal_H,
\ee
where $\Hcal_L$ is the truncated subspace spanned by ``low'' operators with $\De \leq \Dmax$, and $\Hcal_H$ is created by the remaining ``high'' operators. The full invariant mass operator $M^2$ thus takes the schematic form
\be
M^2 = \begin{pmatrix} \Mcal_{LL} & \Mcal_{LH} \\ \Mcal_{HL} & \Mcal_{HH} \end{pmatrix}.
\ee
The matrix $\Mcal_{LL}$, which only acts on the space $\Hcal_L$, corresponds to the truncated version of $M^2$ we diagonalize to obtain the approximate mass eigenstates at a given $\Dmax$. However, there are clearly corrections to this approximation due to the remaining matrix elements.

To understand these corrections more concretely, let's write the true mass eigenstates as
\be
|\mu_i\> = |\mu_i\>_L + |\mu_i\>_H,
\ee
where $|\mu_i\>_{L,H} \in \Hcal_{L,H}$. The \emph{exact} eigenvalue equation can then be rewritten solely in terms of operators acting on the truncated space $\Hcal_L$,
\be
\Big( \Mcal_{LL} - \Mcal_{LH}(\Mcal_{HH} - \mu_i^2)^{-1}\Mcal_{HL} \Big)|\mu_i\>_L = \mu_i^2 |\mu_i\>_L.
\ee
By only diagonalizing the truncated matrix $\Mcal_{LL}$, we've therefore neglected the correction
\be
\de \Mcal \equiv \Mcal_{LH}(\Mcal_{HH} - \mu_i^2)^{-1}\Mcal_{HL}.
\ee
The rate of convergence for conformal truncation is thus set by the asymptotic behavior of $\de \Mcal$ as $\Dmax \ra \infty$. This correction also gives rise to an effective cutoff on our IR resolution, $\LambdaIR$, as we cannot accurately reproduce eigenvalues below the scale set by $\de\Mcal$.

\begin{figure}[t!]
\begin{center}
\includegraphics[width=1\textwidth]{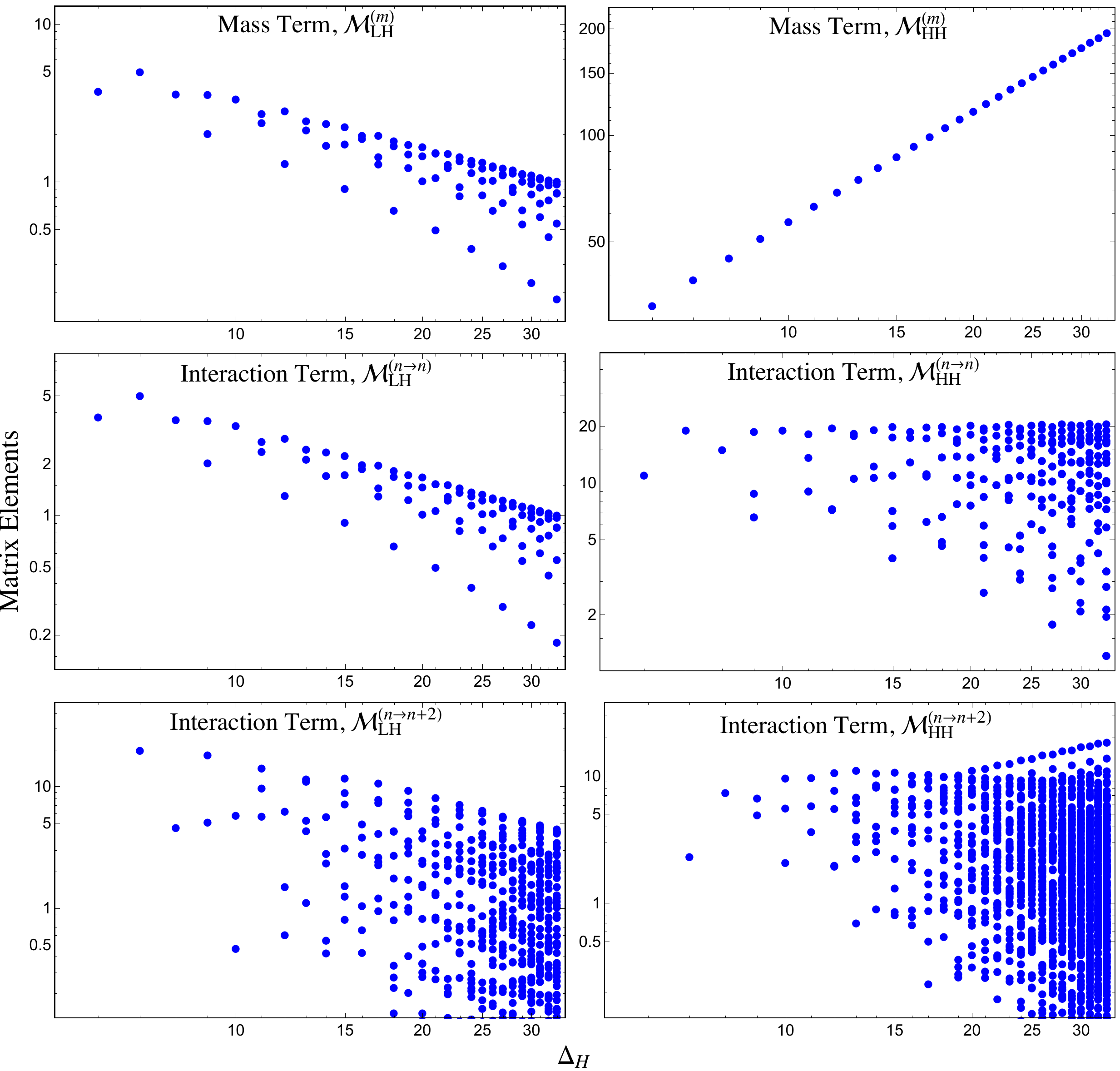}
\caption{Three-particle Casimir eigenstate matrix elements, with overall factors of $m^2$ and $\fr{\lambda}{4\pi}$ removed, as a function of the larger of the two operator scaling dimensions, $\De_H$, for the mass term (top), $n \ra n$ interaction (middle) and $n \ra n+2$ interaction (bottom). Left: matrix elements involving the lowest-dimension operator ($\De_L = 3)$. Right: matrix elements where both operators have dimension $\De_H$.}
\label{fig:MatrixElements} 
\end{center}
\end{figure}

While this correction technically depends on the exact eigenvalues, we're specifically interested in low-mass states. We therefore expect the matrix elements $\Mcal_{HH}$ to be large compared to $\mu_i^2$, which suggests we can approximate the correction as
\be
\de \Mcal \approx \Mcal_{LH}\Mcal_{HH}^{-1}\Mcal_{HL}.
\label{eq:Scaling}
\ee
Given this approximation, we can obtain a rough estimate of the IR cutoff by studying the overall magnitude of matrix elements at the edge of our truncation, involving operators with dimension $\De_H \sim \Dmax$.

Recall that for $\phi^4$ theory, there are three contributions to the Hamiltonian matrix: the mass term, the $n\ra n$ interaction term, and the $n\ra n+2$ interaction term. Figure~\ref{fig:MatrixElements} shows how the individual matrix elements for these three contributions vary with $\Delta_H$ for the case of $n=3$ particles (the other particle sectors are similar). The plots on the left correspond to the `LH' matrix elements, where we have chosen the light state to be the lowest three-particle state, with $\Delta_L=3$, while the plots on the right correspond to `HH' matrix elements. 

From these plots, we can roughly read off the dependence of the largest matrix elements on $\Delta_H$. For the mass matrices, we find
\be
\Mcal^{(m)}_{LH} \sim \fr{1}{\sqrt{\De_H}}, \hspace{10mm} \Mcal^{(m)}_{HH} \sim \De_H.
\label{eq:MassMatrixElements}
\ee
Based on eq.~(\ref{eq:Scaling}), we can use this asymptotic behavior to estimate the IR cutoff, 
\be
\LambdaIR^2 \sim \fr{|\Mcal_{LH}|^2}{\Mcal_{HH}} \sim \fr{m^2}{\Dmax^2}.
\ee
This estimate matches our free field theory results in section~\ref{sec:SanityChecks}, as the corrections to the three-particle threshold (as well as the other $n$-particle thresholds) approximately vanish as $1/\Dmax^2$. 

For the interaction matrices, the $\Mcal_{LH}$ terms also decrease as $\De_H \ra \infty$, though the $n \ra n+2$ matrix elements appear to fall off more slowly than the mass term, suggesting that those elements will provide the dominant contribution at large $\Dmax$. The corresponding $\Mcal_{HH}$ elements are either approximately constant ($n\ra n$) or slowly increasing ($n\ra n+2)$, which indicates that they are both subdominant compared to the rapidly growing mass term.

These matrix elements thus explain the observed behavior of the eigenvalue extrapolations in figure~\ref{fig:ExtrapolationExamples}. At weak coupling, the mass term contribution dominates the IR cutoff, such that the corrections scale as $1/\Dmax^2$. As we increase the coupling, the $\phi^4$ $\Mcal_{LH}$ elements begin to contribute more strongly, slowing the rate of convergence and leading to roughly $1/\Dmax$ corrections near the critical point.

More generally, we learn from these results that the linear growth of the mass term \emph{guarantees convergence} in 2D $\phi^4$ theory. Because the matrix elements mixing higher-dimensional operators with our truncated basis all decrease as we increase $\Dmax$, the suppression from the mass term ensures that our IR cutoff must vanish \emph{at least} as quickly as $1/\Dmax$.

%%%%%%%%%%%%%%%%%%%%%%%%%%%%%%%%%%%%%%%%%%%%%%%%%%%%%%%%%%%%%%%%%%%%%%%%%%%%%
%%%%%%%%%%%%%%%%%%%%%%%%%%%%%%%%%%%%%%%%%%%%%%%%%%%%%%%%%%%%%%%%%%%%%%%%%%%%%

\bibliographystyle{utphys}
\bibliography{Bib2DIsing}

\end{document}